\documentclass[final,3p,authoryear,12pt]{elsarticle}
\usepackage[round]{natbib}

\usepackage[colorinlistoftodos,textsize=footnotesize]{todonotes}

\usepackage{siunitx}
\usepackage{amsfonts,amsmath,amssymb,lscape,float,graphicx,tabularx,url,times,theorem,color,natbib,afterpage,rotating,hyperref,xcolor,tikz,moresize,multirow,enumerate,colortbl,mathtools,bm,bbm}
\usepackage[utf8]{inputenc}
\usepackage{mathtools}
\usepackage[caption=true]{subfig}
\usepackage[ruled, vlined]{algorithm2e}
\usepackage{enumitem}

\usepackage[colorinlistoftodos]{todonotes}
\usepackage{wrapfig}
\usepackage{enumerate}
\usepackage{bbold}
\usepackage{bbm}
\usepackage{subfig}
\usepackage{verbatim}
\usepackage{soul}
\usepackage{ulem}

\usepackage{sectsty}
\sectionfont{\large}
\subsectionfont{\normalsize}
\setcitestyle{comma}

\mathtoolsset{showonlyrefs}
\allowdisplaybreaks 

\usepackage{diagbox}

\newcounter{algsubstate}

\usepackage{setspace} 
 \newcommand{\acknowledgements}[1]{%
  \nonumnote{#1}}
  
\definecolor{myblue}{rgb}{0.8,0.8,1}
\definecolor{myred}{rgb}{1,0.8,0.8}
\definecolor{mygreen}{rgb}{0.8,1,0.8}
\definecolor{mygrey}{rgb}{220,220,220}

\usepackage[bottom]{footmisc}
\usepackage[scaled]{helvet}

\DeclareMathAlphabet{\xcal}{OMS}{cmsy}{m}{n}

\definecolor{dbblue}{RGB}{10,65,155}				
\definecolor{dbred}{RGB}{215,0,50}					
\definecolor{blue}{RGB}{0,113.9850,188.9550} 			
\definecolor{red}{RGB}{216.7500,82.8750,24.9900} 		
\definecolor{green}{RGB}{118.8300,171.8700,47.9400} 	
\definecolor{grey}{RGB}{110,110,110}				
\definecolor{lgrey}{RGB}{210,210,210}				
\definecolor{c1}{RGB}{0,113.9850,188.9550}			
\definecolor{c2}{RGB}{216.7500,82.8750,24.9900}		
\definecolor{c3}{RGB}{236.8950,176.9700,31.8750}		
\definecolor{c4}{RGB}{125.9700,46.9200,141.7800}		
\definecolor{c5}{RGB}{118.8300,171.8700,47.9400}		
\definecolor{c6}{RGB}{76.7550,189.9750,237.9150}		
\definecolor{c7}{RGB}{161.9250,19.8900,46.9200}		
\definecolor{c14}{RGB}{0,102,102}					

\hypersetup{colorlinks,urlcolor=dbblue,citebordercolor=dbblue,linkbordercolor=dbblue,filecolor=dbblue,filebordercolor=dbblue,citecolor=dbblue,linkcolor=dbblue,colorlinks=true}

\def\E{\mathbb{E}}

\defcitealias{UK7year:2013}{Griffiths et al., 2013}

\makeatletter
\def\ps@pprintTitle{%
  \let\@oddhead\@empty
  \let\@evenhead\@empty
  \def\@oddfoot{\reset@font\hfil\thepage\hfil}
  \let\@evenfoot\@oddfoot
}
\makeatother

\newtheorem{thm}{Theorem}

\newtheorem{prop}{Proposition}

\newtheorem{assume}{Assumption}

\newcommand{\diff}{\mathrm{d}}

\newcommand{\Ac}{\mathcal{A}}

\renewcommand{\E}{\mathbbm{E}}
\newcommand{\R}{\mathbbm{R}}

\graphicspath{{images/}{../images/}}

\begin{document}

\title{\textit{Forthcoming in SIAM Journal on Financial Mathematics} \\~\\ \textbf{Decentralised Finance and Automated Market Making: Predictable Loss and Optimal Liquidity Provision}}

\author[omi,math]{\'{A}lvaro Cartea}
\author[omi]{Fayçal Drissi}
\author[omi,math]{Marcello Monga}

\address[omi]{Oxford-Man Institute of Quantitative Finance, Oxford, UK}
\address[math]{Mathematical Institute, University of Oxford, Oxford, UK}

\date{}
\acknowledgements{We are grateful to Tarek Abou Zeid, Carol Alexander, Álvaro Arroyo, Sam Cohen, Patrick Chang, Mihai Cucuringu, Olivier Guéant, Anthony Ledford, Jonathan Plenk, Andre Rzym, and Leandro Sánchez-Betancourt, for insightful comments. The authors thank the Fintech Dauphine Chair, in partnership with Mazars and Crédit Agricole CIB, for their financial support. We are also grateful to seminar participants at Oxford, the OMI, the Oxford Victoria Seminar, the Oxford Mathematical Institute, EWGCFM, QuantMinds International, the WBS Gillmore Centre Academic Conference, and the Oxford DeFi Research Group. FD is grateful to the Oxford-Man Institute's generosity and hospitality.   MM acknowledges financial support from the EPSRC Centre for Doctoral Training in Mathematics of Random Systems: Analysis, Modelling and Simulation (EP/S023925/1).}

\begin{frontmatter}

\begin{abstract}
Constant product markets with concentrated liquidity (CL) are the most popular type of automated market makers. In this paper, we characterise the continuous-time wealth dynamics of strategic LPs who dynamically adjust their range of liquidity provision in CL pools. Their wealth results from fee income, the value of their holdings in the pool, {and rebalancing costs}. Next, we derive a self-financing and closed-form optimal liquidity provision strategy where the width of the LP's liquidity range is determined by the profitability of the pool (provision fees minus gas fees), the predictable losses (PL) of the LP's position, and concentration risk. Concentration risk refers to the decrease in fee revenue if the marginal exchange rate (akin to the midprice in a limit order book) in the pool exits the LP's range of liquidity. When the drift in the marginal rate is stochastic, we show how to optimally skew the range of liquidity to increase fee revenue and profit from the expected changes in the marginal rate. Finally, we use Uniswap v3 data to show that, on average, LPs have traded at a significant loss, and to show that the out-of-sample performance of our strategy is superior to the historical performance of LPs in the pool we consider. 
\end{abstract}


\end{frontmatter}

\section{Introduction  \label{sec:intro}}

Traditional electronic exchanges are organised around limit order books to clear demand and supply of liquidity. In contrast, the takers and providers of liquidity in  constant function markets (CFMs) interact in liquidity pools; liquidity providers (LPs) deposit their assets in the liquidity pool and liquidity takers (LTs) exchange assets directly with the pool. At present, constant product markets (CPMs) with concentrated liquidity (CL) are the most popular type of CFM, with  Uniswap v3 as a prime example; see \cite{uniswap2021core}. In CPMs with CL, LPs specify the rate intervals (i.e., tick ranges) over which they deposit their assets, and this liquidity is counterparty to trades of LTs when the marginal exchange rate of the pool is within the liquidity range of the LPs. When LPs deposit liquidity, fees paid by LTs accrue and are paid to LPs when they withdraw their assets from the pool. The amount of fees accrued to LPs is proportional to the share of liquidity they hold in each liquidity range of the pool.

Existing research characterises the losses of LPs, but does not offer tools for strategic liquidity provision. In this paper, we study strategic liquidity provision in CPMs with CL. We derive the continuous-time dynamics of the wealth of strategic LPs which consists of the position they hold in the pool (position value),  fee income, and costs from repositioning their liquidity position. The width of the range where the assets are deposited affects the value of the LP's position in the pool; specifically, we show that the predictable loss (PL) incurred by LPs increases as the width of the liquidity range decreases. PL measures the unhedgeable losses of LPs stemming from the depreciation of their holdings in the pool and from the opportunity costs from locking their assets in the pool; see \cite{cartea2023predictable}. PL is similar to loss-versus-rebalancing (LVR) in  \cite{milionis2022automated} which describes the losses in traditional CFMs when LPs hedge their price exposure in an exogenous market.\footnote{\cite{milionis2022automated} introduced LVR in August 2022 as a measure that quantifies the unhedgeable losses of LPs in CFMMs when prices form in a centralised venue. Contemporaneously, \cite{cartea2023predictable} introduced PL in November 2022 as a component in the wealth of strategic LPs that quantifies losses due to the convexity of the trading function and due to opportunity costs in CFMs and in CL markets.} 
Also, we show that fee income is subject to a tradeoff between the width of the LP's liquidity range and the volatility of the marginal rate in the pool. More precisely, CL increases fee revenue when the rate is in the range of the LP, but also increases \textit{concentration risk}. Concentration risk refers to the risk that the LP faces when her position is concentrated in narrow ranges; the LP stops collecting fees when the rate exits the range of her position.

We derive an optimal dynamic strategy to provide liquidity in a CPM with CL. In our model, the LP maximises the expected utility of her terminal wealth, which consists of the accumulated trading fees and the gains and losses from the market making strategy. The dynamic strategy controls the width and the skew of liquidity that targets the marginal exchange rate. For the particular case of log-utility, we obtain the strategy in closed-form and show how the solution balances the  opposing effects between PL and fee collection. When volatility increases, PL increases, so there is an incentive for the LP to widen the range of liquidity provision  to reduce the strategy's exposure to PL. In particular, in the extreme case of very high volatility, the LP must withdraw from the pool because the exposure to PL is too high. Also, when there is an increase in the potential provision fees that the LP may collect because of higher liquidity taking activity, the strategy balances two opposing forces. One, there is an incentive to increase fee collection by concentrating the liquidity of the LP in a tight range around the exchange rate of the pool. Two, there is an incentive to limit the losses due to concentration risk by widening the range of liquidity provision. 
Finally, when the drift of the marginal exchange rate is stochastic (e.g., a predictive signal), the strategy skews the range of liquidity to increase fee revenue, by capturing the LT trading flow, and to increase the position value, by profiting from the expected changes in the marginal rate. 

Finally, we use Uniswap v3 data to motivate our model and to test the performance of the strategy we derive. The LP and LT data are from the pool ETH/USDC (Ethereum and USD coin) between the inception of the pool on $5$ May $2021$ and $18$ August $2022$.  To illustrate the performance of the strategy we use in-sample data to estimate model parameters and out-of-sample data to test the strategy. Our analysis of the historical transactions in Uniswap v3 shows that LPs have traded at a significant loss, on average, in the ETH/USDC pool. We show that the out-of-sample performance of our
strategy is considerably superior to the average LP performance we observe in the ETH/USDC pool.

Early works on AMMs are in \cite{chiu2019blockchain, angeris2021analysis, lipton2021blockchain, capponi2021adoption, engel2021composing,engel2021presentation,angeris2022does}; {see also the recent review \cite{biais2023advances}}. {\cite{capponi2023price} study price formation in AMMs.} Some works in the literature study strategic liquidity provision in CFMs and CPMs with CL. \cite{heimbach2022risks} discuss the tradeoff between risks and returns that LPs face in Uniswap v3, \cite{cartea2023predictable}  study the predictable losses of LPs in a continuous-time setup, \cite{milionis2023automated} study the impact of fees on the profits of arbitrageurs in CFMs,  \cite{fukasawa2023model} study the hedging of the impermanent losses of LPs, and \cite{li2023yield} study the economics of liquidity provision. Closest to our work are the models in \cite{fan2021strategic} and \cite{fan2022differential} which focus on fee revenue and use approximation techniques to obtain dynamic strategies. Finally, there is a growing literature on AMM design for fair competition between LPs and LTs. \cite{goyal2023finding} study an AMM with dynamic trading functions that incorporate beliefs of LPs, \cite{lommers2023:case} study AMMs where the LP's strategy adjusts dynamically to market information, and  \cite{cartea2023automated} generalise CFMs and propose AMM designs where LPs express their beliefs and risk preferences; see also \cite{bergault2024price} and \cite{he2024liquidity}.

Our work is related to the algorithmic trading and optimal market making literature.  Early works on liquidity provision in traditional markets are \cite{ho1983dynamics}, \cite{biais1993price}, and  \cite{avellaneda2008high} with extensions in many directions; see \cite{cartea2014buy, cartea2017algorithmic, gueant2017optimal, bergault2021closed, drissi2022solvability}. We refer the reader to \cite{cartea2015algorithmic}, \cite{gueant2016financial}, and \cite{donnelly2022optimal} for a comprehensive review of algorithmic trading models for takers and makers of liquidity in traditional markets. Also, our work is related to those in  \cite{cartea2018enhancing, barger2019optimal, cartea2020market, donnelly2020optimal, forde2022optimal, bergault2022multi} which implement market signals in algorithmic trading strategies. 

The remainder of the paper proceeds as follows. Section  \ref{sec:CL} describes CL pools. Section \ref{sec:wealthCL} studies the continuous-time dynamics of the wealth of LPs as a result of the position value, the fee revenue, and rebalancing costs. In particular, we use Uniswap v3 data to study the fee revenue component of the LP's wealth and our results motivate the assumptions in our model.  Section \ref{sec:4} introduces our liquidity provision model and uses stochastic control to derive a closed-form optimal strategy. Next, we study how the strategy controls the width and the skew of the liquidity range as a function of the pool's profitability, PL, concentration risk, and the drift in the marginal rate. Finally, Section  \ref{sec:num1} uses  Uniswap v3 data to test the performance of the strategy and showcases its superior performance.

\section{Constant function markets and concentrated liquidity \label{sec:CL}}

\subsection{Constant function markets}

Consider a reference asset $X$ and a risky asset $Y$ which is valued in units of $X.$ Assume there is a pool that makes liquidity for the pair of assets $X$ and $Y$, and denote by $Z$ the marginal exchange rate of asset $Y$ in units of asset $X$ in the pool. In a CFM that charges a fee $\tau$ proportional to trade size, the trading function $f\left(q^X, q^Y\right)  = \kappa^2$ links the state of the pool before and after a trade is executed, where $q^X$ and $q^Y$ are the quantities of asset $X$ and $Y$ that constitute the \textit{reserves} in the pool, $\kappa$ is the depth of the pool, and $f$ is increasing in both arguments. We write $f\left(q^X, q^Y\right) = \kappa^2$ as $q^X = \varphi_\kappa\left(q^Y\right)$ for an appropriate decreasing \textit{level function} $\varphi_\kappa$.

We denote the execution rate for a traded quantity $\pm y$ by $\tilde Z\left(\pm y\right)$, where $y\ge 0$. When an LT buys $y$ of asset $Y$, she pays $x = y \times \tilde Z\left(y\right)$ of asset $X$, where 
\begin{equation}\label{eq:cfm_lt_cond_1}
f\left(q^X + (1-\tau)\,x, q^Y - y\right) = \kappa^2 \quad \implies \quad  \tilde Z(y) = \frac{\varphi_\kappa\left(q^Y-y\right) - \varphi_\kappa(q^Y)}{\left(1-\tau\right)\,y}\,.
\end{equation}
Similarly, when an LT sells $y$ of asset $Y$, she receives $x = y \times \tilde Z\left(-y\right)$ of asset $X$, where
\begin{equation}\label{eq:cfm_lt_cond_2}
f\left(q^X + x, q^Y +  (1-\tau)\,y\right) = \kappa^2 \quad \implies \quad  \tilde Z(-y) = \frac{\varphi_\kappa\left(q^Y\right) - \varphi_\kappa\left(q^Y+\left(1-\tau\right)\,y\right)}{y}\,.
\end{equation}

In CL markets, the marginal exchange rate  is $Z = -\varphi_\kappa'(q^Y)$, which is the price of an infinitesimal trade when fees are zero, i.e., when $y\rightarrow 0$ and $\tau = 0$.\footnote{When fees are not zero, the exchange rates for infinitesimal trades are $\lim_{y\rightarrow 0} \tilde Z(y) = - \frac{1}{1-\tau}\, \varphi'_\kappa(y)= \frac{1}{1-\tau}\,Z $ and $\lim_{y\rightarrow 0} \tilde Z(-y) = - \left(1-\tau\right) \varphi'_\kappa(y) = \left(1-\tau\right)\,Z$; see \eqref{eq:cfm_lt_cond_1} and \eqref{eq:cfm_lt_cond_2}.} In traditional CPMs such as Uniswap v2, the trading function is $f\left(q^X, q^Y\right) = q^X\times q^Y$, so the level function is $\varphi_\kappa(q^Y) = \kappa^2/q^Y$ and the marginal exchange rate is $Z = q^X / q^Y.$ 

Liquidity provision operations in CPMs do not impact the marginal rate, so when an LP deposits the quantities $x$ and $y$ of assets $X$ and $Y$, the condition $Z = q^X/q^Y = (q^X+ x)/(q^Y+ y)$ must be satisfied; see \cite{cartea2022decentralised, cartea2023predictable}.

\subsection{Concentrated liquidity markets}

This paper focuses on liquidity provision in CPMs with CL. In CPMs with CL, LPs specify a range of rates $(Z^\ell, Z^u]$ in which their assets can be counterparty to liquidity taking trades. Here, $Z^\ell$ and $Z^u$ take values in a finite set $\{Z^1,\dots,Z^{N}\}$, the elements of the set are called ticks, and the range $(Z^i, Z^{i+1}]$ between two consecutive ticks is a \textit{tick range} which represents the smallest possible liquidity range; see \cite{drissi2023models}  for a description of the mechanics of CL.\footnote{In traditional limit order books, a tick is the smallest price increment.}

The assets that the LP deposits in a range  $(Z^\ell, Z^u]$ provide the liquidity that supports marginal rate movements between $Z^\ell$ and $Z^u$. The quantities $x$ and $y$ that the LP provides verify the key formulae
\begin{align}
\label{eq:kappatilde_def}
\begin{cases}
x=0\quad\quad\quad\quad\quad\  \ \ \, \quad \quad\quad\quad\textrm{and}\quad y=\tilde{\kappa}\left(\left(Z^{\ell}\right)^{-1/2}-\left(Z^{u}\right)^{-1/2}\right) & \textrm{if}\ \ Z\leq Z^{\ell},\\
x=\tilde{\kappa}\left(Z^{1/2}-\left(Z^{\ell}\right)^{1/2}\right)\quad\ \ \ \ \,\textrm{and}\quad y=\tilde{\kappa}\left(Z^{-1/2}-\left(Z^{u}\right)^{-1/2}\right) & \textrm{if}\ \ Z^{\ell}<Z\leq Z^{u},\\
x=\tilde{\kappa}\left(\left(Z^{u}\right)^{1/2}-\left(Z^{\ell}\right)^{1/2}\right)\ \ \ \,\textrm{and}\quad y=0 & \textrm{if}\ \ Z>Z^{u}\,,
\end{cases}
\end{align}
where $\tilde \kappa$ is the depth of the LP's liquidity in the pool.  The depth  $\tilde \kappa$  is specified by the LP and it remains constant unless the LP provides additional liquidity or withdraws her liquidity. When the rate $Z$ changes, the equations in \eqref{eq:kappatilde_def} and the prevailing marginal rate $Z$  determine the holdings of the LP in the pool, in particular, they determine the quantities of each asset received by the LP when she withdraws her liquidity.

Within each tick range, the constant product formulae \eqref{eq:cfm_lt_cond_1}--\eqref{eq:cfm_lt_cond_2} determine the dynamics of the marginal rate, where the depth $\kappa$ is the total depth of liquidity in that tick range. To obtain the total depth in a tick range, one sums the depths of the individual liquidity positions in the same tick range; see \cite{drissi2023models}. When a liquidity taking trade is large, so the marginal rate crosses the boundary of a tick range, the pool executes two separate trades with potentially different depths for the constant product formula.

In CPMs with CL, the proportional fee $\tau$ charged by the pool to LTs is distributed among LPs. More precisely, if an LP's liquidity position with depth $\tilde \kappa$ is in a tick range where the total depth of liquidity is $\kappa$, then for every liquidity taking trade that pays an amount $p$ of fees,\footnote{{The proportional fee $\tau$ represents a fixed percentage applied to the trade size to calculate the fee $p$ paid by an LT. Thus, the amount $p$ is  expressed in terms of the reference asset $X$.}} the LP with liquidity $\tilde \kappa$ earns the amount
\begin{align}
\label{eq:remuneration_LP}
\tilde  p = \frac{\tilde{\kappa}}{ \kappa}\, p\,\mathbbm{1}_{Z^{\ell} < Z \leq Z^{u}}\,.
\end{align}
Thus, the larger is the position depth $\tilde \kappa$, the higher is the proportion of fees that the LP earns.\footnote{For instance, if the LP is the only provider of liquidity in the range $(Z^\ell, Z^u]$ then $\kappa = \tilde \kappa$, so the LP collects all the fees in that range.} The equations in \eqref{eq:kappatilde_def} imply that for equal wealth, narrow liquidity ranges increase the value of $\tilde \kappa.$ However, as the liquidity range of the LP decreases, the concentration risk increases.

\section{The wealth of liquidity providers in CL pools \label{sec:wealthCL}}

In this section, we consider a strategic LP who dynamically tracks the marginal rate $Z$. In our model, the LP's position is self-financed throughout an investment window $[0, T]$, so the LP repeatedly withdraws her liquidity and collects the accumulated fees, then uses her wealth, i.e., the collected fees and the assets she withdraws, to deposit liquidity in a new range. In the remainder of this work, we work in a filtered probability space $\left(\Omega, \mathcal F, \mathbbm P; \mathbbm F = (\mathcal F_t)_{t \in [0,T]} \right)$ that satisfies the usual conditions, where $\mathbbm F$ is the natural filtration generated by the collection of observable stochastic processes that we define below.

We assume that the marginal exchange rate  in the pool $\left(Z_t\right)_{t\in [0,T]}$ is driven by a stochastic drift $\left(\mu_t\right)_{t\in [0,T]}$ and we write
\begin{align}
\label{eq:dynZ}
   \diff Z_t = \mu_t \, Z_t\, \diff t + \sigma\, Z_t\, \diff W_t\,,  
\end{align}
where the volatility parameter $\sigma$ is a nonnegative constant and $(W_t)_{t \in [0,T]}$ is a standard Brownian motion independent of $\mu$. We assume that $\mu$ is càdlàg with finite fourth moment, i.e., $\mathbbm E\left[\mu_t^4\right]<+\infty$ for $ 0\leq t\leq T$. The LP observes and uses $\mu$ to optimise her liquidity positions and improve trading performance.

Consider an LP with initial wealth $\tilde x_0,$ in units of $X,$ and an investment horizon $[0, T],$ with $T>0.$ At time $t=0\,,$ she deposits  quantities $\left(x_0, y_0\right)$ in the range $\left(Z^\ell_0, Z^u_0\right]$, so the initial depth of her position is $\tilde \kappa_0,$ and the value of her initial position, marked-to-market in units of $X$, is $\tilde x_0 = x_0 + y_0 \, Z_0.$   The dynamics of the LP's wealth consist of the value  of the liquidity position in the pool $\left(\alpha_t\right)_{t \in [0, T]}$, the fee revenue $\left(p_t\right)_{t \in [0, T]}$, and the rebalancing costs $\left(c_t\right)_{t \in [0, T]}$.  We introduce the wealth process $(\tilde x_t = \alpha_t +  p_t + c_t)_{t \in [0,T]}$, which we mark-to-market in units of the reference asset $X$, with $\tilde x_0 > 0$ known.  At any time $t,$ the LP uses her wealth $\tilde x_t$ to provide liquidity. Next, Subsection  \ref{sec:positionvalue} studies the dynamics of the LP's position $\alpha$ in the pool, Subsection \ref{sec:feerevenue} studies the dynamics of the LP's fee revenue $p$, and Subsection \ref{sec:costs} studies the dynamics of the rebalancing costs $c$.

\subsection{Position value \label{sec:positionvalue}}

In this section, we focus our analysis on the \textit{position value} $\alpha$.  Throughout the investment window $[0, T]$, the holdings $\left(x_t, y_t\right)_{t\in[0,T]}$ of the LP change because the marginal rate $Z$ changes and because she continuously adjusts her liquidity range around $Z$. More precisely, to make markets optimally, the LP controls the values of $\left(\delta_t^\ell\right)_{t\in[0,T]}$ and  $\left(\delta_t^u\right)_{t\in[0,T]}$ which determine the dynamic liquidity provision boundaries $\left(Z_t^\ell\right)_{t\in[0,T]}$ and $\left(Z_t^u\right)_{t\in[0,T]}$ as follows:
\begin{align}
\label{eq:ZlZucontrol}
\begin{cases}
\left(Z_{t}^{u}\right)^{1/2}= & Z_{t}^{1/2}/\left(1-\delta_{t}^{u}/2\right),\\
\left(Z_{t}^{\ell}\right)^{1/2}= & Z_{t}^{1/2}\left(1-\delta_{t}^{\ell}/2\right),
\end{cases}
\end{align}
where $\delta^\ell \in \left(-\infty, 2\right]$, $\delta^u \in \left[-\infty, 2\right)$, and $\delta^\ell\,\delta^u/2<\delta^\ell+\delta^u$ because $0 \leq Z^\ell < Z^u < \infty.$ We define $\delta^\ell$ and $\delta^u$ in \eqref{eq:ZlZucontrol} in terms of $\sqrt Z$ to simplify and linearise the CL constant product formulae; see \cite{cartea2023predictable} for more details.

In practice, the LP earns fees when the rate $Z_t$ is in the LP's liquidity range $(Z^\ell_t, Z^u_t]$, so  $\delta^\ell \in \left(0, 2\right]$, $\delta^u \in \left[0, 2\right)$, and $\delta^\ell\,\delta^u/2<\delta^\ell+\delta^u$.\footnote{Recall that in practice $Z^\ell$ and $Z^u$ take values in a finite set of ticks, so $\delta^\ell$ and $\delta^u$ also take values in a finite set. In the liquidity provision problem of Section \ref{sec:4}, we use stochastic control techniques to derive an optimal strategy where the controls $\delta^\ell$ and $\delta^u$ are continuous, so we round the values of $Z^\ell$ and $Z^u$ to the nearest ticks in the performance analysis of Section \ref{sec:num1}.} Below, Section \ref{sec:4} considers a problem where the controls are not constrained, and values $\delta^\ell \notin \left(0, 2\right]$, $\delta^u \notin \left[0, 2\right)$ are those where liquidity provision is unprofitable.

In the remainder of this paper we define the \textit{spread} $\delta_t$ of the LP's position  as
\begin{align}
\label{eq:pos_spread}
\delta_t = \delta_t^u + \delta_t^\ell\,,
\end{align}
and we  consider admissible strategies $\delta$ that are $\R\textrm{-valued}$ and such that $\int_{0}^{T}\delta_{t}^{-4}\,\diff t<\infty,$ almost surely; see Section \ref{sec:4} for more details. For small position spreads, we use the first-order Taylor expansion to write the approximation $$\left(Z_t^u - Z^\ell_t\right)\Big/ Z_t = \left(1-\delta_{t}^{u}/2\right)^{-2}-(1-\delta_{t}^{\ell}/2)^{2} \approx \delta_t.$$

We  assume  that the marginal rate process $\left(Z_t\right)_{t\in [0,T]}$ follows the dynamics \eqref{eq:dynZ}. \cite{cartea2023predictable} show that the holdings in assets $X$ and $Y$ in the pool for an LP who follows an arbitrary strategy $\left(Z_t^\ell, Z_t^u\right)$ are given by
\begin{equation}\label{eq:holdingspool}
x_t =    \frac{\delta_{t}^{\ell}}{\delta_{t}^{\ell}+\delta_{t}^{u}}\,\alpha_{t} \quad \text{and} \quad  y_t = \frac{\delta_{t}^{u}}{Z_t\left( \delta_{t}^{\ell}+\delta_{t}^{u}\right) }\,\alpha_{t}\,,
\end{equation}
so the value $\left(\alpha_t\right)_{t \in [0, T]}$ of her position follows the dynamics
\begin{align}\label{eq:dynAlphaFirst}
\diff\alpha_{t}=&\,\,\tilde x_{t}\,\left(\frac{1}{\delta_{t}^{\ell}+\delta_{t}^{u\,}}\right)\left(-\frac{\sigma^{2}}{2}\,\diff t+\mu_t\,\delta_{t}^{u}\,\diff t+\sigma\,\delta_{t}^{u}\,\diff W_{t}\right)\nonumber \\
=&\,\, \diff \text{PL}_t + \tilde x_{t}\,\left(\frac{1}{\delta_{t}^{\ell}+\delta_{t}^{u\,}}\right)\left(\mu_t\,\delta_{t}^{u}\,\diff t+\sigma\,\delta_{t}^{u}\,\diff W_{t}\right)\,,
\end{align}
where the predictable and negative component $\text{PL}_t = -\frac{\sigma^2}{2}\, \int_0^t \frac{ \tilde x_s}{ \delta_s} \, \diff s $ is the PL of the LP's position which scales with the volatility of the marginal rate. PL is related to LVR (see \cite{milionis2022automated}) which decomposes the loss of LPs in traditional CFMs into a hedgeable market risk and an unhedgeable component due to profits made by arbitrageurs. The dynamics in \eqref{eq:dynAlphaFirst} show that a larger position spread $\delta$ reduces PL and the overall risk of the LP's position in the pool; see \cite{cartea2023predictable} for more details. 

For a fixed value of the spread $\delta_t = \delta^\ell_t + \delta^u_t$, the dynamics in \eqref{eq:dynAlphaFirst} show that if $\mu_t \ge 0$, then the LP increases her expected wealth by increasing the value $\delta^u$, i.e., by skewing her range of liquidity to the right. However, note that the quadratic variation of the LP's position value is $\diff \langle \alpha, \alpha \rangle_t = \tilde x_t^2 \, \sigma^2 \left(\delta_t^u / \delta_t\right)^2\, \diff t\,,$ so skewing the range to the right also increases the variance of the LP's position. On the other hand, if $\mu \le 0$, then the LP reduces her losses by decreasing the value of $\delta^u$ or equivalently increasing the value of $\delta^\ell$, i.e., by skewing her range of liquidity to the left. Thus, the LP uses the expected changes in the marginal rate to skew the range of liquidity and to increase her expected terminal wealth.

\subsection{Fee income \label{sec:feerevenue}}

\subsubsection{Fee income: pool fee rate}

The dynamics of the fee income in our model of Section \ref{sec:4} uses a fixed depth $\kappa$ and assumes that the pool generates fee income for all LPs at an instantaneous \textit{pool fee rate} $\pi$; clearly, these fees are paid by LTs who interact with the pool, see \eqref{eq:cfm_lt_cond_1}--\eqref{eq:cfm_lt_cond_2}.  The value of $\pi$ represents the instantaneous profitability of the pool, akin to the size of market orders and their arrival rate in LOBs.

{In contrast to the proportional fee $\tau$, which represents the fixed percentage applied to the trade sizes of LTs to compute the fees paid to the pool, the pool fee rate $\pi$ denotes the instantaneous profitability of the pool  computed as the percentage of total fees paid by LTs relative to the size of the pool. Thus, the pool fee rate $\pi$ depends on the intensity of the liquidity taking flow, the size of the pool, and the proportional fee $\tau$. Below, we also consider the fee revenue $p$ of one LP who maximises her wealth by choosing an optimal spread for her liquidity position.}

\begin{figure}[!htb]\centering
\includegraphics[width=0.45\textwidth]{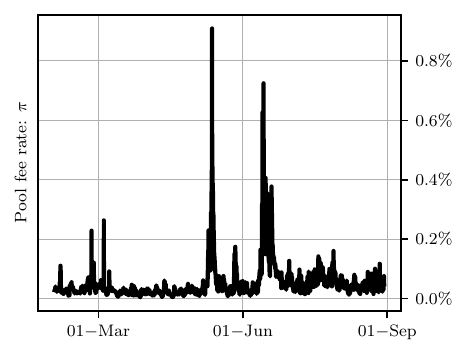}
\includegraphics[width=0.45\textwidth]{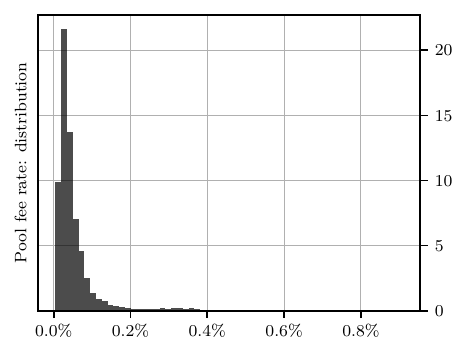}\\
\caption{ 
Estimated pool fee rate from February to August 2022 in the ETH/USDC pool. For any time $t,$ the pool fee rate is the total fee income, as a percentage of the total pool size, paid by LTs on the period $[t-1\text{ day}, t].$ The pool size at time $t$ is $2 \, \kappa \, Z_t^{1/2}$ in terms of asset $X$, where $Z_t$ is the active rate in the pool at time $t.$ Left panel: historical values of the pool fee rate. Right panel: distribution of the pool fee rate. }
\label{fig:DOLP_feerate}
\end{figure}

To analyse the  dynamics of the pool fee rate $\pi,$ we use historical LT transactions in Uniswap v3 as a measure of activity and to estimate the total fee income generated by the pool; \ref{sec:appx:data} describes the data and Table \ref{table:datadescr} provides descriptive statistics. Figure \ref{fig:DOLP_feerate} shows the estimated  fee rate $\pi$ in the ETH/USDC pool. For any time $t,$ we use\footnote{Pools have different fee rates in Uniswap v3: $0.01\%$, $0.05\%,$ and $0.1\%.$} $$\pi_t = 0.05\% \, \frac{V_t}{2\,\kappa\, Z_t^{1/2}}  \,,$$ where $V_t$ is the volume of LT transactions the day before $t,$  $2\,\kappa\, Z_t^{1/2}$ is the pool value in terms of asset $X$ at time $t,$ and $0.05 \%$ is the fixed fee of the pool.\footnote{The value of a CPM pool is given by the active pool depth. In particular, the size  of a pool with quantity $x$ of asset $X$ and quantity $y$ of asset $Y$ is $x+Z\,y = 2\,x =  2\,\kappa\, Z^{1/2}$ units of $X$ because $x\, y =\kappa^2 \ \text{and} \ Z= x/ y \ \text{ so } x^2 = \kappa^2 \, Z. $} Figure \ref{fig:DOLP_feerate} suggests that the pool fee rate $\pi$ generated by liquidity taking activity in the pool is stochastic and mean reverting. Here, we assume that $\pi$ is independent of the rate $Z$ over the time scales we consider; Table \ref{table:corrPiZ} shows that the pool fee rate is weakly correlated to the  rate $Z$ at different sampling frequencies, especially for the higher frequencies we consider in our numerical tests. In Section \ref{sec:4}, the pool fee rate $\pi$ follows Cox-Ingersoll-Ross-type dynamics.

\begin{table}[h]
\footnotesize
\begin{center}
\begin{tabular}{c  r  r  r  r} 
\hline 
& $\Delta t = 1$ minute & $\Delta t = 5$ minutes & $\Delta t = 1$ hour & $\Delta t = 1$ day \\ [0.5ex] 
\hline
Correlation & $-2.1 \%\ \ $ & $-2.4 \%$ & $-2.6 \%$ & $-10.9 \%$ \\ 
{Two-tailed p-value} & {$2.0\%\ \ $} & {$8.9\%$} & {$8.2\%$} & {$16.7\%$} \\ 
{$R^2$ regression} & {$0.01\%$} & {$1.1\%$} & {$2.9\%$} & {$5.0\%$} \\ 
\hline 
\end{tabular}
\end{center}
\caption {{First row:} correlation of the returns of the rate $Z$ and the fee rate $\pi$, i.e., $\left(Z_{t+\Delta t} - Z_t\right) /  Z_t$ and { $\left(\pi_{t+\Delta t} - \pi_t\right)  /  \pi_t$ }for $\Delta t = 1$ minute, five minutes, one hour, and one day, using data of the ETH/USDC pool  between 5 May 2021 and 18 August 2022. {Second row: two-tailed p-value of the t-test. Final row: $R$-squared of regression of the pool fee rate's returns against the marginal exchange rate's returns.}}
\label{table:corrPiZ}
\end{table}


\subsubsection{{Fee income: spread and concentration risk} \label{sec:concentrationcost}}

\begin{figure}[!h]
\centering
\includegraphics{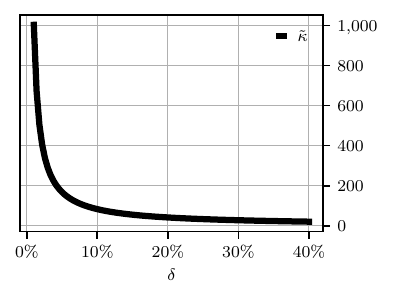}\\
\caption{Value of the depth $\tilde \kappa$ of the LP's position in the pool as a function of the spread $\delta.$ The spread is in percentage of the marginal exchange rate; recall that  $\left(Z_t^u - Z^\ell_t\right) / Z_t  \approx \delta_t.$}\label{fig:SIMPLE1}
\end{figure}

In the three cases of \eqref{eq:kappatilde_def}, increasing the spread reduces the depth $\tilde \kappa$ of the LP's position.  Recall that the LP fee income is proportional to $\tilde \kappa / \kappa,$ where $\kappa$ is the pool depth. Thus, decreasing the value of $\tilde \kappa$  potentially reduces LP fee income. Figure \ref{fig:SIMPLE1} shows the value of $\tilde \kappa$ as a function of the spread $\delta$.

However, although narrow ranges increase the potential fee income, they also increase concentration risk; a wide spread (i.e., a lower value of the depth $\tilde \kappa$)  decreases fee income per LT transaction but reaps earnings from a larger number of LT transactions because the position is active for longer periods of time (i.e., it takes longer, on average, for $Z$ to exit the LP's range). Thus, the LP must strike a balance between maximising the depth $\tilde \kappa$ around the rate and minimising the concentration risk, which depends on the volatility of the rate $Z.$

In our model, the LP continuously adjusts her position around the current rate $Z,$ so we write the continuous-time dynamics of  \eqref{eq:remuneration_LP}, conditional on the rate not exiting the LP's range, as
\begin{align}
\label{eq:feeDynamics00}
\diff p_{t}= \underbrace{ \left(\tilde{\kappa}_{t} \, / \, \kappa\right)}_{\text{Position depth}} \ \underbrace{\pi_t }_{\text{Fee rate}} \ \underbrace{\,2 \,\kappa \,Z_{t}^{1/2} }_{\text{Pool size}} \, \diff t\,,
\end{align}
where $(\tilde \kappa_t)_{t \in [0,T]}$ models the depth of the LP's position and $p$ is the LP's fee income for providing  liquidity with depth $\tilde\kappa$ in the pool. The fee income is proportional to the pool size, i.e., proportional to $2\,\kappa\, Z_t^{1/2}.$ Next, use the second equation in \eqref{eq:kappatilde_def} and  equations \eqref{eq:ZlZucontrol}--\eqref{eq:holdingspool} to write the dynamics of the LP's position depth $\tilde \kappa_t$ as
\begin{align}
\label{eq:dynKappaTildeDis0}
\tilde{\kappa}_{t}=2\,\tilde{x}_{t}\,\left(\frac{1}{\delta_{t}^{\ell}+\delta_{t}^{u}}\right)\,Z_{t}^{-1/2}\,,
\end{align}
so the dynamics in \eqref{eq:feeDynamics00} become
\begin{align}
\label{eq:feeDynamics0}
\diff p_{t} = \left(\frac{4}{\delta_{t}^{\ell}+\delta_{t}^{u}}\right)\, \pi_t\, \tilde{x}_{t}\, \diff t\,.
\end{align}

In practice, {there is latency in the market and the LP cannot reposition her liquidity  position and rebalance her assets continuously.  Thus,  
the LP faces concentration risk in between the times she repositions her liquidity; narrow spreads generate less fee income because the rate $Z$ may exit the range of the LP's liquidity, especially in volatile markets.}

{
To model the losses due to concentration risk in the continuous-time dynamics  \eqref{eq:feeDynamics0} of the LP's fee revenue, we introduce an instantaneous concentration cost that reduces the fees collected by the LP as a function of the spread;} the concentration cost increases (decreases) when the spread narrows (widens). We modify the dynamics of the fees collected by the LP  in \eqref{eq:feeDynamics0}  as follows
\begin{align}
\label{eq:feedyn_0}
\diff p_{t} = \left(\frac{4}{\delta_{t}^{\ell}+\delta_{t}^{u}}\right)\, \pi_t\, \tilde{x}_{t}\, \diff t\, - \gamma\,\left(\frac{1}{\delta_t^{\ell}+\delta_t^{u}}\right)^{2}\,\tilde{x}_t\,\diff t\,,
\end{align}
where $\gamma>0$ is the concentration cost parameter and $\tilde x_t$ is the wealth invested by the LP in the pool at time $t.$ 

{
To justify the form of the concentration cost, we study the realised fee revenue  in the ETH/USDC pool as a function of the spread $\delta$ in \eqref{eq:pos_spread} for rebalancing frequencies $m=1$ minute and $m=5$ minutes. We denote by $\widehat  p_{\delta,m}$ the average realised fee revenue earned by an LP in the ETH/USDC pool who provides liquidity with wealth $\tilde x = 1$, in a range with spread $\delta$, around the marginal rate $Z$ throughout a time window of length $m$. The left panel of Figure \ref{fig:gamma_1_5min} shows that the value of $\delta$ which maximises the fee revenue $\widehat  p_{\delta,m}$ is strictly positive for both values of $m$ that we consider. In particular, narrow ranges reduce fee revenue due to concentration risk.

The form of the fee revenue dynamics \eqref{eq:feedyn_0} is a second order Taylor approximation that captures the specific shape of the fee revenue in CL markets; see left panel of Figure \ref{fig:gamma_1_5min}. The LP uses the regression model
\begin{equation}
    \delta^2 \, \widehat p_{\delta,m} = 4\,\pi\,m\,\delta - \gamma \, m\,,
\end{equation} 
which is based on the dynamics \eqref{eq:feedyn_0}, to estimate the concentration cost parameter $\gamma$. 
The right panel of Figure \ref{fig:gamma_1_5min} shows that $\delta^2 \, \widehat p_{\delta,m}$ is affine in $\delta$ and that the estimated concentration cost parameter $\gamma$ depends on the rebalancing frequency of the LP. In particular, the figure shows that the dynamics \eqref{eq:feedyn_0} and the second order approximation are suitable to describe the realised fee revenue in CL markets. The performance study of Section \ref{sec:num1} uses this methodology to set the value of the concentration cost parameter $\gamma$.

\begin{figure}[!h]\centering
\includegraphics{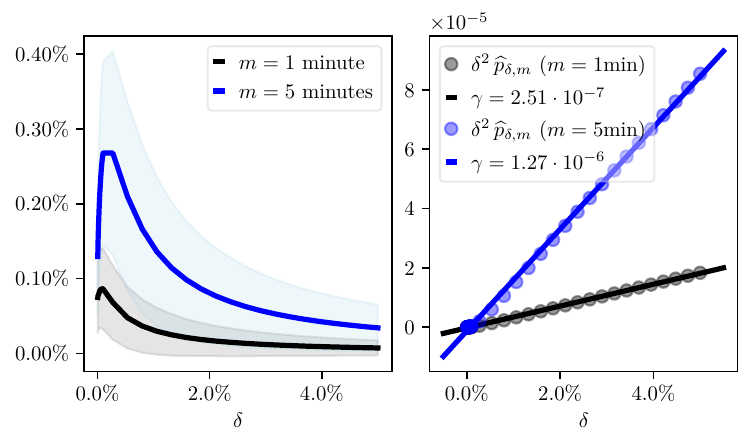}\\
\caption{{Left panel: mean and standard deviation of the fee revenue of hypothetical liquidity positions as a function of the position spread $\delta$ for the rebalancing frequencies $m=1$ minute and $m=5$ minutes. The fee revenue of each position considers a wealth $\tilde x=1$ and uses historical LT transactions of the ETH/USDC pool of Uniswap v3; see data description in \ref{sec:appx:data}. Right panel: $\delta^2\,\widehat p_{\delta,m}$ as a function of the position spread $\delta$ for the rebalancing frequencies $m=1$ minute and $m=5$ minutes. The concentration cost parameter is estimated as $-\iota / m$ where $\iota$ is the intercept of the regression of $\delta^2\,\widehat p_{\delta,m}$ on $\delta.$}  }\label{fig:gamma_1_5min}
\end{figure}

}

\subsubsection{Fee income: drift and asymmetry}

The stochastic drift $\mu$ indicates the future expected changes of the marginal exchange rate in the pool. In practice, the LP may use a predictive signal so that $\mu$ represents the belief that the LP holds over the future marginal rate. For an LP who maximises fee revenue, it is natural to consider asymmetric liquidity positions that capture the liquidity taking flow. We define the \textit{asymmetry} of a position as 
\begin{align}
\label{eq:asymmetryrhot}
\rho_t = \delta^u_t  /  \left(\delta^u_t + \delta^\ell_t \right) = \delta^u_t  / \delta_t\, ,
\end{align}
where $\delta^u_t$ and $\delta^\ell_t$ are defined in \eqref{eq:ZlZucontrol}. In one extreme, when the asymmetry $\rho \to 0,$ then $Z^u \to Z$ and  the position consists of only asset $X,$ and in the other extreme, when $\rho \to 1,$ then $Z^\ell \to Z$ and the position consists of only asset $Y.$ 

\begin{figure}[!h]\centering
\includegraphics{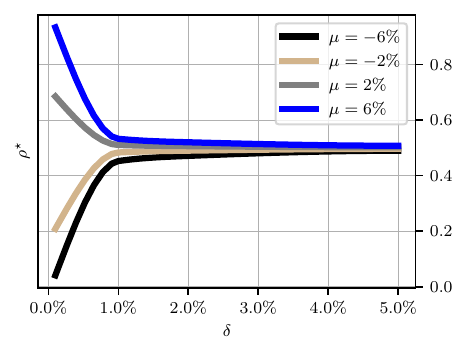}
\caption{Optimal position asymmetry $\rho^\star$ in \eqref{eq:asymmetryrhot} as a function of the spread  $\delta$ of the position, for multiple values of the drift $\mu.$ {For each historical value of the drift in our data, we compute the performance of hypothetical liquidity positions with asymmetry values in the range $[0, 1]$ and position spreads in the range $[0.05\%, 5\%].$} The  asymmetry $\rho^\star$ in the figure is the value of $\rho$ that maximises the average fee income for each value of the pair $\left(\delta, \mu\right)$. {The drift is computed as the mean of $5$-minute returns during one hour of trading, and the performance of the liquidity positions is computed as the total fee revenue if the position is held for an hour. } }\label{fig:DOLP5}
\end{figure}

{In the remainder of this work, the asymmetry of the LP's position is a function of the observed drift:
\begin{align}
\label{eq:relation_rho_mu}
\rho_t = \rho\left(\delta_t, \mu_t\right) = \frac12 + \frac{\mu_t}{\delta_t} = \frac12 + \frac{\mu_t}{\delta_t^u + \delta_t^\ell} \,, \quad \forall t \in [0, T]\,.
\end{align}
The asymmetry \eqref{eq:relation_rho_mu} adapts the skew of the position to the expected drift of the marginal rate. When $\mu=0$, the position is symmetric around the marginal rate and $\rho_t = 1/2$, so $\delta_t^\ell = \delta_t^u.$ When $\mu>0,$ the position is skewed to the right (i.e., $\delta_t^u > \delta^\ell_t$) to capture more LT trades and fee revenue, and similarly when $\mu<0,$ the position is skewed to the left (i.e., $\delta_t^u < \delta^\ell_t$). Also, when $\mu>0$ and the position is skewed to the right according to \eqref{eq:relation_rho_mu}, the proportion of asset $Y$ increases and the position profits from rate appreciation, and when the position is skewed to the left, the proportion of asset $Y$ decreases and the position is protected from rate depreciation.

Optimal liquidity provision is the dynamic choice of  $\delta^{u}$ and $\delta^{\ell}$ over a trading window, or equivalently, the dynamic choice of  $\delta$ and $\rho$. Our model assumes that the LP fixes the asymmetry $\rho$ of her position at time $t$ according to \eqref{eq:relation_rho_mu}, so we reduce the trading problem to a one-dimensional dynamic optimisation problem, which significantly simplifies calculations. 

To further justify the form of the asymmetry }\eqref{eq:relation_rho_mu}, we use Uniswap v3 data to study how the asymmetry and the width of the LP's range of liquidity relate to fee revenue. First, we estimate the realised drift $\mu$ in the pool ETH/USDC over a rolling window of $T=5$ minutes.\footnote{The values of the drift in this section are normalised to reflect daily estimates. In particular, we use $\mu = \tilde \mu \, / \, \Delta t$ where $\tilde{\mu}$ is the average of the observed log returns and  $\Delta t$ is the observed average LT trading frequency.} Next, for any time $t$, the fee income for different positions of the LP's liquidity range is computed for various values of the spread $\delta$ and for various values of the asymmetry $\rho.$ For each value of the realised drift $\mu$ during the investment horizon, and for each fixed value of the spread $\delta,$ we record the asymmetry that maximises fee income. Figure \ref{fig:DOLP5} shows the optimal (on average) asymmetry $\rho$ as a function of the spread  $\delta$ of the position for multiple values of the realised drift $\mu.$ 

Figure \ref{fig:DOLP5} suggests that there exists a preferred  asymmetry of the position for a given value of the spread $\delta$ and a given  value of the drift  $\mu.$ First, for all values of the spread $\delta$, the LP skews her position to the right when the drift is positive ($\rho^\star > 0.5$) and she skews her position to the left when the drift is negative ($\rho^\star < 0.5$). Second, for narrow spreads, the liquidity position requires more asymmetry than for large spreads when the drift is not zero. 

In our liquidity provision model of Section \ref{sec:4}, the LP  holds a belief over the future exchange rate throughout the investment window and controls the spread $\delta=\delta^u + \delta^\ell$  of her position. Thus, she strategically chooses the asymmetry of her position as a function of $\delta$ and $\mu.$  We approximate the relationship exhibited in Figure  \ref{fig:DOLP5} with the asymmetry function \eqref{eq:relation_rho_mu}.


{

\subsection{Rebalancing costs and gas fees \label{sec:costs}}

Our model considers a strategic LP who continuously repositions her liquidity position in the pool which requires rebalancing of the LP's assets.  Specifically, repositioning typically leads to different holdings \eqref{eq:holdingspool} in the pool, in which case the LP trades one asset for the other in the pool or in another trading venue. However, rebalancing assets to reposition liquidity is costly. 

Let $\left(c_t\right)_{t\in[0,T]}$ denote the cost of rebalancing in terms of asset $X$. Similar to \cite{fan2021strategic} and \cite{fan2022differential}, we model rebalancing costs as proportional to the quantity $y_t$ of asset $Y$ that the LP deposits in the pool.  At any time $t$, we assume that the LP uses all her wealth $\tilde x_t$ when repositioning her liquidity position, so we use \eqref{eq:holdingspool} to write
\begin{equation}\label{eq:rebalancing_cost_dyns}
c_t = -\zeta\,y_t\,Z_t = -\zeta\,\frac{\delta_{t}^{u}}{\delta_{t}^{\ell}+\delta_{t}^{u} }\,\tilde{x}_{t}\,, \quad c_0=0\,,
\end{equation}
where $\zeta>0$ is a constant that models the execution costs.

Transactions sent to the pool also bear gas fees. Gas fees are a flat fee paid to the blockchain and do not depend on the size of the LP's transaction; see \cite{li2023yield} for more details on the impact of gas fees on liquidity provision. Thus, gas fees scale with the frequency at which the LP sends transactions to the pool. Our model considers continuous trading, so gas fees do not influence the optimisation problem, but should be considered when the strategy is implemented.}

In the next section, we derive an optimal liquidity provision strategy, and prove that the profitability of liquidity provision  is subject to a tradeoff between fee revenue, PL, and concentration risk. 

\section{Optimal liquidity provision in CL pools \label{sec:4}}

\subsection{The problem}

Consider an LP who provides liquidity in a CPM with CL throughout the investment window $[0, T]$. We work on the filtered probability space $\left(\Omega, \mathcal F, \mathbbm P; \mathbbm F = (\mathcal F_t)_{t \in [0,T]} \right)$ where $\mathcal F_t$ is the natural filtration generated by the collection $\left(Z, \mu, \pi\right)$.

The dynamics of the LP's wealth consist of the fees earned, the position value, and {the rebalancing costs}.  Similar to Section \ref{sec:positionvalue}, we denote the wealth process of the LP by $(\tilde x_t = \alpha_t +  p_t + c_t)_{t \in [0,T]},$ with $\tilde x_0 > 0$ known, where $\alpha$ is the value of the LP's position, $p$ is the fee revenue{, and $c$ is the rebalancing cost}. At any time $t,$ the LP uses her wealth $\tilde x_t$ to provide  liquidity. {Now, use \eqref{eq:dynAlphaFirst}, \eqref{eq:feedyn_0}, and \eqref{eq:rebalancing_cost_dyns} to write the dynamics of the LP's wealth as
\begin{align}
\label{eq:temp_cash_dyn}
\diff\tilde{x}_{t}=\tilde{x}_{t}\left(\frac{1}{\delta_{t}^{\ell}+\delta_{t}^{u}}\right)\left[\left(4\,\pi_{t}-\frac{\sigma^{2}}{2}+\left(\mu_{t}-\zeta\,\right)\,\delta_{t}^{u}\right)\diff t+\sigma\,\delta_{t}^{u}\,\diff W_{t}\right]-\gamma\,\left(\frac{1}{\delta_{t}^{\ell}+\delta_{t}^{u}}\right)^{2}\tilde{x}\,\diff t\,,\ \ 
\end{align}
where $\gamma \geq 0$ is the concentration cost parameter, see Subsection \ref{sec:feerevenue}, and $\pi$ follows the dynamics defined below in \eqref{eq:dyn_pi}. To simplify notation, we set $\zeta=0$; our results hold when replacing $\mu$ by $\mu-\zeta$.}

Next, use $\delta_t = \delta^u_t + \delta_t^\ell$ and $\delta^u_t \, / \, \delta_t = \rho\left(\mu_t, \delta_t\right)$ in \eqref{eq:temp_cash_dyn} to write the dynamics of the LP's wealth as
\begin{align}
\label{eq:dynxtilde}
\diff\tilde{x}_{t}&=\frac{1}{\delta_{t}}\,\left(4\,\pi_{t}-\frac{\sigma^{2}}{2}\right)\,\tilde{x}_{t}\,\diff t+\mu_t\,\rho\left(\delta_{t},\mu_t\right)\,\tilde{x}_{t}\,\diff t+\sigma\,\rho\left(\delta_{t},\mu_t\right)\,\tilde{x}_{t}\,\diff W_{t}-\frac{\gamma}{\delta_{t}^{2}}\,\tilde{x}_{t}\,\diff t\,.
\end{align}

{Next, following the discussions of Section \ref{sec:feerevenue}, we denote by $\eta$ the \textit{profitability threshold} and we assume that the pool fee rate $\pi$ follows the dynamics}
\begin{align}
\label{eq:dyn_pi}
\diff(\pi_{t} - \eta_t)&=\Gamma\,\left(\overline{\pi}+\eta_t-\pi_{t}\right)\diff t+\psi\,\sqrt{\pi_{t}-\eta_t}\,\diff B_{t}\,,
\end{align}
where  $\Gamma>0$ denotes the mean reversion speed, $\overline{\pi} > 0$ is the long-term mean of $\left(\pi_t - \eta_t\right)_{t\in[0,T]}$, $\psi>0$ is a non-negative volatility parameter, $\left(B_t\right)_{t\in[0,T]}$ is a Brownian motion independent of $\left(W_t\right)_{t\in[0,T]},$ and $\pi_0-\eta_0>0$ is known. 
{To solve the LP's optimal liquidity provision problem, we introduce the following assumption.
\begin{assume}\label{assumption:eta}
The profitability {threshold} $\eta$ in the dynamics  \eqref{eq:dyn_pi} is given by
\begin{equation}\label{eq:eta stochastic}
\eta_t=\frac{\sigma^{2}}{8}-\frac{\mu_t}{4}\left(\mu_t-\frac{\sigma^{2}}{2}\right)+\frac{\varepsilon}{4}\,.
\end{equation}
\end{assume}}

From Assumption \ref{assumption:eta} and the CIR dynamics \eqref{eq:dyn_pi} it follows that
\begin{align}\label{assumption:1}
    \pi_{t} - \eta_t \geq 0 \implies 4\,\pi_{t}-\frac{\sigma^{2}}{2}+\mu_t\left(\mu_t-\frac{\sigma^{2}}{2}\right)\ge\varepsilon>0\,, \quad \forall t \in [0, T]\,.
\end{align}
{Assumption \ref{assumption:eta} and condition \eqref{assumption:1}} ensure that the spread $\delta$ of the optimal strategy is {admissible.  Financially, the inequality in \eqref{assumption:1} is a profitability condition $\pi_t \ge \eta_t$ that 
guarantees that the LP's fee income $\pi$ is greater than the PL faced by the LP, adjusted by the drift in the marginal rate.} The profitability condition \eqref{assumption:1} is further discussed in Sections \ref{sec:4:discussion} and \ref{sec:modelling_assumptions}.

Finally, note that the dynamics in \eqref{eq:dynZ} and \eqref{eq:dyn_pi} imply that the LP also observes $W,$ $B,$ and the {profitability threshold} $\eta$ is determined by $\mu$, so the LP observes all the stochastic processes of this problem.

\subsection{The optimal strategy}

The LP controls the spread  $\delta$ of her position to maximise the expected utility of her terminal wealth in units of $X$. {To define the set of admissible strategies $\mathcal{A}$, note that if the LP assumes that the asymmetry function $\rho$ is that in  \eqref{eq:relation_rho_mu}, then for each $\delta$, we need
\begin{equation}\label{eq:rho_condition}
    \int_{0}^{T} \rho\left(\delta_{t},\mu_t\right)^2 \,\diff t<\infty  \quad \mathbbm{P}\textrm{--a.s.}\,.
\end{equation}
Observe that 
\begin{equation}
    \begin{split}
        \int_{0}^{T} \rho\left(\delta_{t},\mu_t\right)^2 \,\diff t \,= \,& \int_{0}^{T} \left(\frac12 + \frac{\mu_t}{\delta_t}\right)^2 \,\diff t\,
        \leq \, \frac T4 + \frac{1}{2}\,\int_{0}^{T} \mu_t^4 \,\diff t+\frac{1}{2}\,\int_{0}^{T} \frac{1}{ \delta_t^4} \,\diff t\,.
    \end{split}
\end{equation}
Thus, to satisfy \eqref{eq:rho_condition} and to ensure that the control problem below is well-posed, we define the set of admissible strategies
\begin{align}
\label{def:admissibleset_t}
\mathcal{A}_{t}=\left\{ (\delta_{s})_{s\in[t,T]},\ \R\textrm{-valued},\ \mathbbm{F}\textrm{-adapted, and }\int_{t}^{T}\frac{1}{\delta_{s}^{4}}\,\diff s<+\infty \ \mathbbm{P}\textrm{--a.s.}\right\} \, ,
\end{align}
where $\mathcal A := \mathcal A_0.$ 
}

Let $\delta\in\Ac$. The performance criterion of the LP is a function $u^{\delta}\colon[0,T] \times \R^4 \rightarrow \R$ given by
\begin{align}
\label{eq:perfcriteria}
    u^{\delta}(t,\tilde{x},z, \pi,\mu)=\mathbbm{E}_{t,\tilde{x},z, \pi,\mu}\left[U\left(\tilde{x}_{T}^{\delta}\right)\right]\, ,
\end{align}
where $U$ is a concave utility function, and the value function $u:[0,T] \times \R^4 \rightarrow \R$ of the LP is
\begin{equation}
\label{eq:valuefunc}
    u(t,\tilde{x},z,\pi,\mu)=\underset{\delta\in\mathcal{A}_t}{\sup} \, u^{\delta}(t,\tilde{x},z,\pi,\mu)\, .
\end{equation}

The following results solve the optimal liquidity provision model when the LP assumes a general stochastic drift $\mu$ and her preferences are given by a logarithmic utility function.

\begin{prop}
\label{prop:1}
Assume the asymmetry function $\rho$ is as  in  \eqref{eq:relation_rho_mu} and that $U(x) = \log(x).$ Then,
\begin{align}
\label{eq:hjbsol}
w\left(t,\tilde{x},z,\pi,\mu\right)=&\log\left(\tilde{x}\right)+\left(\pi-\eta\right)^{2}\int_{t}^{T}\E_{t,\mu}\left[\frac{8}{2\,\gamma+\mu_{s}^{2}\,\sigma^{2}}\right]\exp\left(-2\,\Gamma\left(s-t\right)\right)\,\diff s\\&+\left(\pi-\eta\right)\left(2\,\Gamma\,\overline{\pi}+\psi^{2}\right)\int_{t}^{T}\E_{t,\mu}\left[C\left(s,\mu_{s}\right)\right]\exp\left(-\Gamma\left(s-t\right)\right)\,\diff s\\&-\left(\pi-\eta\right)\int_{t}^{T}\E_{t,\mu}\left[\frac{4\,\varepsilon}{2\,\gamma+\sigma^{2}\,\mu_{s}^{2}}\right]\exp\left(-\Gamma\left(s-t\right)\right)\,\diff s\\&+\int_{t}^{T}\left(\Gamma\,\overline{\pi}\,\E_{t,\mu}\left[E\left(s,\mu_{s}\right)\right]+\psi^{2}\,\E_{t,\mu}\left[\eta_{s}\,C\left(s,\mu_{s}\right)\right]\right)\diff s
\\
&-\frac{1}{2}\,\int_{t}^{T}\left(\E_{t,\mu}\left[\frac{\varepsilon^{2}}{2\,\gamma+\sigma^{2}\,\mu_{s}^{2}}+\mu_{s}\right]\right)\diff s-\pi\frac{\sigma^{2}}{8}\left(T-t\right)\,
\end{align}
solves the HJB equation associated with problem \eqref{eq:valuefunc}. Here, $\eta_s = \frac{\sigma^{2}}{8}-\frac{\mu_s}{4}\left(\mu_s-\frac{\sigma^{2}}{2}\right)+\frac{\varepsilon}{4}$ for $s\ge t\,,$ $\eta_t = \eta$, and  $\E_{t,\mu}$ represents expectation conditioned on $\mu_t = \mu$, and $$C\left(t,\mu\right)=\E_{t,\mu}\left[\,\int_{t}^{T}\frac{8}{2\,\gamma+\mu_{s}^{2}\,\sigma^{2}}\exp\left(-2\,\Gamma\left(s-t\right)\right)\,\diff s\right]\,,$$ and $$E\left(t,\mu\right)=\E_{t,\mu}\left[\,\int_{t}^{T}\left(\left(2\,\Gamma\,\overline{\pi}+\psi^{2}\right)C\left(s,\mu\right)+\frac{4\,\varepsilon}{2\,\gamma+\sigma^{2}\,\mu_{s}^{2}}\right)\exp\left(-\Gamma\left(s-t\right)\right)\,\diff s\right]\,.$$

\end{prop}

\noindent
For a proof, see \ref{sec:proofs:hjb1}.

\begin{thm}
\label{thm:verif}
Let {Assumption \ref{assumption:eta} hold and assume that } the asymmetry function $\rho$ is as in \eqref{eq:relation_rho_mu} and that $U(x)= \log(x).$ Then, the solution in Proposition \ref{prop:1} is the unique solution to the optimal control problem \eqref{eq:valuefunc}, and the optimal spread $\left(\delta_s\right)_{s\in[t,T]} \in \mathcal A_t$ is given by
\begin{align}
\label{eq:optimalspeed}
\delta_{s}^{\star}=\frac{2\,\gamma+\mu_s^{2}\,\sigma^{2}}{4\,\pi_{s}-\frac{\sigma^{2}}{2}+\mu_s\left(\mu_s-\frac{\sigma^{2}}{2}\right)}=\frac{2\,\gamma+\mu_s^{2}\,\sigma^{2}}{4\left(\pi_{s}-\eta_s\right)+\varepsilon}\,,
\end{align}
where $\eta_s = \frac{\sigma^{2}}{8}-\frac{\mu_s}{4}\left(\mu_s-\frac{\sigma^{2}}{2}\right)+\frac{\varepsilon}{4}\,.$
\end{thm}

\noindent
For a proof, see \ref{sec:proofs:hjb2}.

\subsection{Discussion: profitability, PL, and concentration risk \label{sec:4:discussion}}

{In this section, we study the strategy 
when $\mu = 0\,,$ in which case the position is symmetric, so $\rho = 1/2$ and $\delta^u_t = \delta^\ell_t = \delta_t / 2$ and the optimal spread \eqref{eq:optimalspeed} becomes}\footnote{The position range is approximately symmetric around the position rate $Z$ because $\delta^u_t = \delta^\ell_t$ does not imply that $Z-Z^\ell = Z^u-Z;$ see \eqref{eq:ZlZucontrol}. However, for small values of $\delta^\ell$ and $\delta^u,$ one can write the approximation $Z-Z^\ell \approx Z^u-Z$, in which case the position is symmetric around the rate $Z$.}
\begin{align}
\label{eq:optimalspreadsul_symm}
\delta_{t}^{\ell\,\star}=\delta_{t}^{u\,\star}=\frac{2\,\gamma}{8\,\pi_{t}-\sigma^{2}}\,\quad \implies \quad \delta_{t}^{\star}=\frac{4\,\gamma}{8\,\pi_{t}-\sigma^{2}},
\end{align}
so the inequality in \eqref{assumption:1} becomes \begin{align}\label{assumption:1_symm}
    4\,\pi_{t}-\frac{\sigma^{2}}{2}\ge\varepsilon>0\,, \quad \forall t \in [0, T]\,.
\end{align}

The inequality in \eqref{assumption:1_symm} guarantees that the optimal control \eqref{eq:optimalspreadsul_symm} does not explode, and ensures that fee income is large enough for LP activity to be profitable. In particular, it ensures that $\pi > \sigma^2/8 + \varepsilon.$ When $\varepsilon \to 0$, i.e., $\sigma^2/4 \to \pi,$ the spread $\delta \to +\infty\,.$ However, we require that the spread $\delta = \delta^u + \delta^\ell \le 4$, so the conditions $\delta^\ell \leq 2$ and $\delta^u \leq 2$ become
\begin{align}
\label{eq:cond0}
\frac{\gamma}{4\,\pi-\frac{\sigma^{2}}{2}}\leq2 \implies \pi-\frac{\gamma}{8}\geq\frac{\sigma^{2}}{8} \,. 
\end{align}

When $\delta^\ell = \delta^u = 2,$ the LP provides liquidity in the maximum range $(0\,, +\infty),$ so the depth of her liquidity position $\tilde \kappa$ is minimal, the PL is minimal, and the LP's position is equivalent to providing liquidity in CPMs without CL; see \cite{cartea2023predictable} for more details. In that case, the dynamics of PL in \eqref{eq:dynAlphaFirst} are $$\diff \text{PL}_t = - \frac{\sigma^2 }{8} \,  \alpha_t\, \diff t\,,$$ so $\sigma^2 / 8$ is the lowest rate at which the LP's assets can depreciate due to PL.

On the other hand, when $\delta\leq4$, the depreciation rate of the LP's position value in \eqref{eq:dynAlphaFirst} is higher {than $\sigma^2/8$}. In particular, if $\delta = \delta^\text{tick}$, where $\delta^\text{tick}$ is the spread of a liquidity position concentrated within a single tick range, then the depth of the LP's liquidity position $\tilde \kappa$ is maximal and PL is maximal with dynamics $$\diff \text{PL}_t = - \frac{\sigma^2 }{2 \, \delta^\text{tick}} \,  \alpha_t\, \diff t\,,$$ so $\sigma^2 / 2 \, \delta^\text{tick}$ is the highest rate at which the LP's assets can depreciate due to PL.

LPs should track the profitability of the pools they consider and check if the expected fee revenue covers PL before considering depositing their assets in the pool. When $\mu=0,$ we propose that LPs use $\sigma^2 / 8$ as a {lower bound} rule-of-thumb for the  pool's rate of profitability because $\sigma^2 / 8$ is the lowest rate of depreciation of their wealth in the pool.

\begin{figure}[!h]\centering
\includegraphics{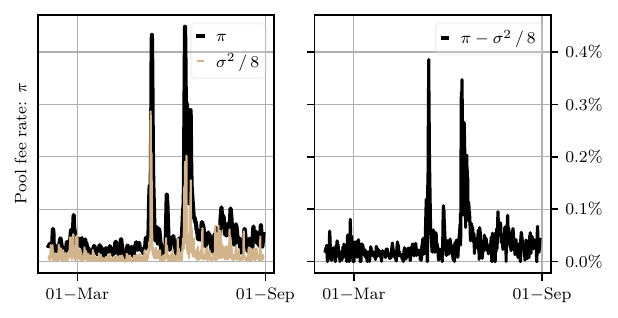}\\
\caption{Estimated pool fee rate from February to August 2022 in the ETH/USDC pool. For any time $t$, the pool fee rate is the total fee income, as a percentage of the total pool size, paid by LTs in the period $[t-1\text{ day}, t].$ The pool size at time $t$ is $2 \, \kappa \, Z_t^{1/2}$ where $Z_t$ is the active rate in the pool at time $t.$ }\label{fig:DOLP_feerate_vol}
\end{figure}

Condition \eqref{eq:cond0} ensures that the profitability $\pi - \gamma / 8,$ which is the pool fee rate adjusted by the concentration cost, is higher than the depreciation rate of the LP's assets in the pool. Thus, the condition imposes a  minimum profitability level of the pool, so LP activity is viable. An optimal control $\delta^\star > 4$ indicates non-viable LP activity because fees are not enough to compensate for the PL borne by the LP. Figure \ref{fig:DOLP_feerate_vol} shows the estimated pool fee rate and the estimated depreciation rate in the ETH/USDC pool (from January to August 2022).  In particular, the CIR model captures the dynamics of $\pi_t - \sigma^2 / 8.$

\begin{figure}[!h]
\centering 
\includegraphics{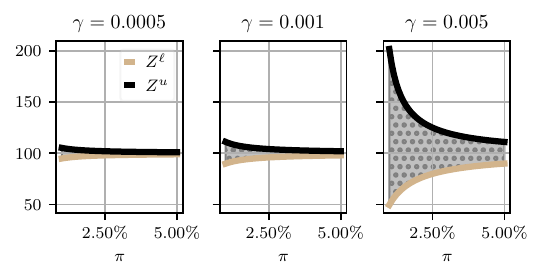}\\
\caption{Optimal LP position range $\left(Z^\ell, Z^u\right]$ as a function of the pool fee rate $\pi$ for different values of the concentration cost parameter $\gamma,$ when $Z = 100,$ $\sigma = 0.02,$ and $\mu=0.$}\label{fig:OLP_dyn0}
\end{figure}

\begin{figure}[!h]
\centering 
\includegraphics{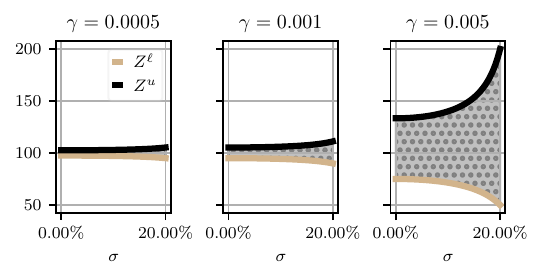}\\
\caption{Optimal LP position range $\left(Z^\ell, Z^u\right]$ as a function of the volatility $\sigma$ for different values of the concentration cost parameter $\gamma,$ when $Z = 100,$ $\pi = 0.02,$ and $\mu=0.$}\label{fig:OLP_dyn2}
\end{figure}

Next, we study the dependence of the optimal spread on the value of the concentration cost coefficient $\gamma,$ the fee rate $\pi,$ and the volatility $\sigma$. The concentration cost coefficient $\gamma$ scales  the spread linearly in \eqref{eq:optimalspreadsul_symm}. Recall that the cost term penalises small spreads because there is a risk that the rate will exit the LP's range. Thus, large values of $\gamma$ generate large values of the spread. Figure \ref{fig:OLP_dyn0} shows the optimal spread as a function of the pool fee rate $\pi$. Large potential fee income pushes the strategy towards targeting more closely the marginal rate $Z$ to profit from fees. Lastly, Figure \ref{fig:OLP_dyn2} shows that the optimal spread increases as the volatility of the rate $Z$ increases.

Finally, the optimal spread does not depend on time or on the terminal date $T.$ The LP marks-to-market her wealth in units of $X,$ but does not penalise her holdings in asset $Y$. In particular, the LP's performance criterion does not include a running penalty or a final liquidation penalty (to turn assets into cash or into the reference asset). For example, if at the end of the trading window the holdings in asset $Y$ must be exchanged for $X$, then the optimal strategy would skew, throughout the trading horizon,  the liquidity range  to convert holdings in $Y$ into $X$ through LT activity.\footnote{In LOBs, one usually assumes that final inventory is liquidated with adverse price impact and that there is a running inventory penalty, thus market making strategies in LOBs depend on the terminal date $T$.  }

\subsection{Discussion: drift and position skew}

In this section, we study how the strategy depends on the stochastic drift $\mu$. Use $\delta_t = \delta^\ell_t + \delta^u_t$ and $\rho\left(\delta_t, \mu_t\right) = \delta^u_t / \delta_t$ to write the two ends of the optimal spread as
\begin{align}
\label{eq:optimalspreadsul}
\delta_{t}^{u \, \star}=\frac{2\,\gamma+\mu_t^{2}\,\sigma^{2}}{8\,\pi_{t}-\sigma^{2}+2\,\mu_t\left(\mu_t-\frac{\sigma^{2}}{2}\right)}+\mu_t \quad \text{and} \quad \delta_{t}^{\ell\, \star}=\frac{2\,\gamma+\mu_t^{2}\,\sigma^{2}}{8\,\pi_{t}-\sigma^{2}+2\,\mu_t\left(\mu_t-\frac{\sigma^{2}}{2}\right)}-\mu_t\,.
\end{align}

The inequality in \eqref{assumption:1} guarantees that the optimal control in \eqref{eq:optimalspreadsul} does not explode and ensures that fee income is large enough for LP activity to be profitable. The profitability condition in \eqref{eq:cond0} becomes
\begin{equation}
\label{eq:cond1}
\pi_{t}-\frac{\gamma}{8}\ge\frac{\sigma^{2}}{8}\left(\frac{\mu_t^{2}\,}{2}+1\right)-\frac{\mu_t}{4}\left(\mu_t-\frac{\sigma^{2}}{2}\right)\,,
\end{equation}
so LPs that assume a stochastic drift in the dynamics of the exchange rate $Z$ should use this simplified measure of the depreciation rate due to PL as a rule-of-thumb before considering depositing their assets in the pool.

\begin{figure}[!h]
\centering 
\includegraphics{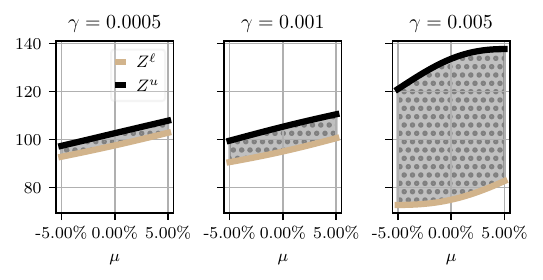}\\
\caption{Optimal LP position range $\left(Z^\ell, Z^u\right]$ as a function of the drift $\mu$ for different values of the concentration cost parameter $\gamma,$ when $Z = 100,$ $\pi = 0.02,$ and $\sigma=0.02.$}\label{fig:OLP_dyn1}
\end{figure}

Next, we study the dependence of the optimal spread on the value of the drift $\mu$.  First, recall that the controls in \eqref{eq:optimalspreadsul} must obey the inequalities\footnote{The admissible set of controls is not restricted to these ranges. However, values outside these range cannot be implemented in practice.}
$$0<\delta^{\ell}_t\leq2 \quad \text{and}\quad 0\leq\delta^{u}_t<2\,,$$
because $0 \leq Z^\ell < Z^u < \infty$ and $Z_t \in (Z^\ell_t, Z^u_t]$, which together with \eqref{eq:pos_spread} implies $0 \leq\delta_t\leq 4$. Next, the asymmetry function satisfies 
\begin{equation}\label{eq:rho}
    0<\rho\left(\delta_t, \mu\right) = \frac{\delta^u_t}{\delta_t}<1\,,
\end{equation}
which implies
\begin{equation}
\label{eq:double_rho}
0\leq\rho \left(\delta_t, \mu\right)\,\delta_t<2\quad \text{and}\quad 0\leq\left(1-\rho \left(\delta_t, \mu\right)\right)\,\delta_t<2\,.
\end{equation}
Now, use \eqref{eq:relation_rho_mu} and \eqref{eq:double_rho} to write
\begin{equation}
\label{eq:double_rho_two}
0\leq\left(\frac{1}{2}+\frac{\mu}{\delta_t}\right)\,\delta_t<2\quad \text{and}\quad
0\leq\left(\frac{1}{2}-\frac{\mu}{\delta_t}\right)\,\delta_t<2\,.
\end{equation}
Finally, use \eqref{eq:rho}  and \eqref{eq:double_rho_two} to obtain the inequalities
\begin{align}
\label{eq:conditionFin}
2\,\left|\mu\right|\leq\delta_t\leq4-2\,\left|\mu\right|,
\end{align}
so $\mu$ must be in the range $\left[-1, 1\right]$ for the LP to provide liquidity. If $\mu$ is outside this range, concentration risk is too high so the LP must withdraw her holdings from the pool. Recall that the dynamics of $Z$ are geometric and $\mu$ is a percentage drift, so values of $\mu$ outside the range $\left[-1, 1\right]$ are unlikely. Moreover, when $\mu=-1,$ the drift of the exchange rate $Z$ is large and negative, so the optimal range is $(0, Z],$ i.e., the largest possible range to the left of $Z.$ When $\mu=1,$ the drift of the exchange rate $Z$ is large and positive, so the optimal range is $(Z, +\infty),$ which is the largest possible range to the right of $Z.$ Condition \eqref{eq:conditionFin} is always verified when we study the performance of the strategy in the ETH/USDC pool. Figure \ref{fig:OLP_dyn1} shows how the optimal spread adjusts to the value of the drift $\mu$. Finally, note that
\begin{equation}\label{eq:derivative_mu}
    \frac{\partial\delta^{u \, \star}}{\partial\sigma}=\frac{\partial\delta^{\ell \, \star}}{\partial\sigma}=\frac{2\,\mu^{2}\,\sigma\left(4\,\pi-4\,\eta+\varepsilon\right)+4\,\sigma\left(1+\mu\right)\left(2\,\gamma+\mu^{2}\,\sigma^{2}\right)}{\left(4\,\pi-4\,\eta+\varepsilon\right)^{2}}>0\,,\qquad \forall\mu\in[-1,1]\,,
\end{equation}
shows that the optimal range is strictly increasing in the volatility $\sigma$ of the rate $Z,$ which one expects as increased activity  that exposes the position value to more PL, and increases the concentration risk.

\section{Performance of strategy\label{sec:num1}}

\subsection{Methodology}
In this section, we use Uniswap v3 data between 1 January and 18 August 2022 (see data description in \ref{sec:appx:data}) to study the performance of the strategy of Section \ref{sec:4}. We consider execution costs and discuss how gas fees and liquidity taking activity in the pool affect the performance of the strategy.\footnote{In practice, LPs pay gas fees when using the Ethereum network to deposit liquidity, withdraw liquidity, and adjust their holdings. Gas is paid in Ether, the native currency of the Ethereum network, and measures the computational effort of the LP operation; see \cite{cartea2022decentralised, cartea2023execution}.} 

Our strategy in Section \ref{sec:4} is solved in continuous time. In our performance study, we discretise the trading window in one-minute periods and the optimal spread is fixed at the beginning of each time-step. That is, let $t_i$ be the times where the LP interacts with the pool, where $i\in\{1,\dots,N\}$ and $t_{i+1} - t_i = 1$ minute.  For each time $t_i,$ the LP uses the optimal strategy in \eqref{eq:optimalspreadsul} based on information available at time $t_i,$ and she fixes the optimal spread of her position throughout the period $[t_i, t_{i+1});$ recall that the optimal spread is not a function of time.

To determine the optimal spread \eqref{eq:optimalspreadsul} of the LP's position at time $t_i,$ we use in-sample data $[t_i-\text{1 day}, t_i]$ to estimate the parameters. The volatility $\sigma$ of the rate $Z$ is given by the standard deviation of one-minute  log returns of the rate $Z,$ which is multiplied by $\sqrt{1440}$ to obtain a daily estimate. The pool fee rate $\pi_t$ is given by the total fee income generated by the pool during the in-sample period, divided by the pool size $2 \, \kappa \, Z_t^{1/2}$ at time $t,$ where $\kappa$ is the active depth at time $t.$ {We run the linear regression described in Section \ref{sec:concentrationcost} to estimate} the concentration cost parameter, {which is set} to $\gamma = 5 \times 10^{-7}$.  
Finally, prediction of the future marginal rate $Z$ is out of the scope of this work, thus we set $\mu=0$ and $\rho = 0.5.$ 

To compute the LP's performance as a result of changes in the value of her holdings in the pool (position value), and as a result of fee income, we use out-of-sample data $[t_i, t_{i+1}].$ {We use  equation \eqref{eq:dynAlphaFirst} to determine the one-minute out-of-sample changes in the position value}. For fee income, we use  LT transactions in the pool at rates included in the range $\left(Z_t^\ell, Z_t^u \right]$ and equation \eqref{eq:remuneration_LP}. The income from fees accumulates in a separate account in units of $X$ with zero risk-free rate.\footnote{In practice, fees accumulate in both assets $X$ and $Y.$} At the end of the out-of-sample window, the LP withdraws her liquidity and collects the accumulated fees, and we repeat the in-sample estimation and out-of-sample liquidity provision described above. Thus, at times $t_i,$ where $i\in\{1,\dots,N-1\},$ the LP consecutively withdraws and deposits liquidity in different ranges. Between two consecutive operations (i.e., reposition liquidity provision), the LP may need to take liquidity in the pool to adjust her holdings in asset $X$ and $Y.$ In that case, we use results in \cite{cartea2022decentralised} to compute execution costs.\footnote{\cite{cartea2022decentralised} show that execution costs in the pool are a closed-form function of the rate $Z,$ the pool depth $\kappa,$ and the transaction size.} In particular, we consider execution costs when the LP trades asset $Y$ in the pool to adjust her holdings between two consecutive operations. More precisely, we consider that for every quantity $y$ of asset $Y$ bought or sold in the pool, a transaction cost $y \, Z_t^{3/2}/\kappa$ is incurred. We assume that the LP's taking activity does not impact the dynamics of the pool. Finally, we obtain 331,858 individual LP operations from 1 January to 18 August 2022.

\subsection{Benchmark}

\begin{figure}[!h]
\centering
\includegraphics[width=\textwidth]{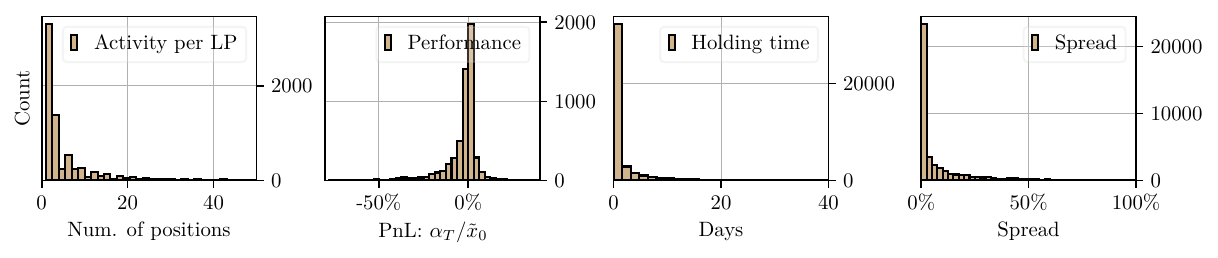}\\
\caption{From left to right: distribution of the number of operations per LP, changes in the holdings value as a percentage of initial wealth, position hold time, and position spread. ETH/USDC pool with selected operations from 5,156 LPs between 5 May 2021 and 18 August 2022.}\label{fig:LP1}
\end{figure}

We compare the performance of our strategy with the performance of LPs in the pool we consider. We select operation pairs that consist of first providing and then withdrawing the same depth of liquidity $\tilde \kappa$ at two different points in time by the same LP.\footnote{In blockchain data, every transaction is associated to a unique wallet address.} The operations that we  select represent approximately $66\%$ of all LP operations. Figure \ref{fig:LP1} shows the distribution of key variables that describe how LPs provide liquidity. The figure shows the distribution of the number of operations per LP, the changes in the position value, the length of time the position is held in the pool, and the position spread. 

Finally, Table \ref{table:dataLPdescr} shows the average and standard deviation of the distributions in Figure \ref{fig:LP1}. Notice that the bulk of liquidity is deposited in small ranges, and positions are held for short periods of time; $20\%$ of LP positions are held for less than five minutes and $30\%$ for less than one hour. Table \ref{table:dataLPdescr} also shows that, on average, the performance of the LP operations in the pool and the period we consider is $-1.49\,\%$ per operation.

{
\begin{table}[h]
\footnotesize{
\begin{center}
\begin{tabular}{c | r  r  r  r  r}
\hline 
 & Number of   & Position value & Fee income & Hold time & Spread \\ [0.5ex]
 & transactions per LP  & performance  ($\alpha_T / \tilde x_0 - 1$) & ($ p_T / \tilde x_0 - 1$) & & \\ [0.5ex]
\hline 
 Average  & $11.5$ & $-1.64\%$& $0.155\%$& $6.1$ days & $18.7\%$\\ [0.5ex]
Standard deviation & $40.2$ & $7.5\%\ \ $ & $0.274\%$ & $22.4$ days &  $43.2\%$\\
\hline 
\end{tabular}
\end{center}
\caption {LP operations statistics in the ETH/USDC pool using operation data of 5,156 different LPs between 5 May 2021 and 18 August 2022. Performance includes transaction fees and excludes gas fees. The position value performance and the fee income are not normalised by the hold time. }
\label{table:dataLPdescr}
}
\hfill
\end{table}}

\subsection{Performance results}
\begin{figure}[!h]
\centering 
\includegraphics{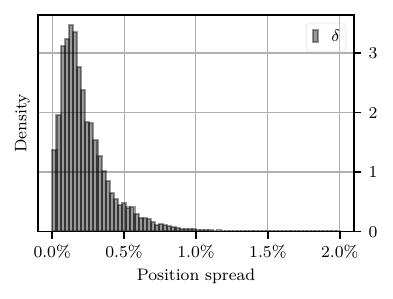}\\
\caption{Distribution of the position spread $\delta.$}\label{fig:DOLP_backtestgammas}
\end{figure}

This subsection focuses on the performance of our strategy when gas fees are zero --- at the end of the section we discuss the profitability of the strategy when gas fees are included. Figure \ref{fig:DOLP_backtestgammas} shows the distribution of the optimal spread \eqref{eq:optimalspreadsul_symm} posted by the LP. The bulk of liquidity is deposited in ranges with a spread  $\delta$ below $1 \%$. Table \ref{table:dataLPdescr} compares the average performance of the components of  the optimal  strategy with  the performance of LP operations observed in the ETH/USDC pool.\footnote{In particular, performance is given for the selected operations shown in Figure \ref{fig:LP1}.} Table \ref{table:dataLPperf} suggests that the position of the LP loses value in the pool (on average) because of PL; however, the fee income would cover the loss, on average, if one assumes that gas fees are zero. Finally, the results show that our optimal strategy significantly improves the PnL of LP activity in the pool and the performance of the assets themselves.

{
\begin{table}[h]
\begin{center}
\begin{footnotesize}
\begin{tabular}{c || r  r  r} 
\hline 
 & Position value performance & Fee income & Total performance \\ 
  &  per operation   &  per operation & per operation  \\ 
    &     &  &  (with transaction costs, \\ 
      &     &  &  without gas fees) \\ 
  [0.5ex]
\hline
Optimal strategy & $-0.015 \%$  & $ 0.0197 \%$ & $0.0047 \% $ \\ [-0.5ex]
 &   $(0.0951 	 \%)$ & $( 0.005 \%)$ & $( 0.02 \%)$ \\ [1ex]
 
Market &  $ -0.0024 \%$  & $ 0.0017 \%$ & $ -0.00067 \% $ \\ [-0.5ex]
 &   $(0.02 \%)$ & $( 0.005 \%)$ & $( 0.02 \%)$ \\ [1ex]

Hold & n.a.   & n.a.  & $-0.00016 \% $ \\ [-0.5ex]
 &    &  & $( 0.08 \%)$ \\ [0.5ex]
\hline
\hline 
\end{tabular}
\end{footnotesize}
\end{center}
\caption {\textbf{Optimal strategy}: Mean and standard deviation of the one-minute  performance of the LP strategy \eqref{eq:optimalspreadsul_symm} and its components. \textbf{Market}: Mean and standard deviation of one-minute performance of LP activity in the ETH/USDC pool using data between 1 January and 18 August 2022.
\textbf{Hold}: Mean and standard deviation of the one-minute  performance of holding the assets. In all cases, the performance includes transaction costs (pool fee and execution cost), but does not include gas fees.}
\label{table:dataLPperf}
\hfill
\end{table}}

The results in Table \ref{table:dataLPperf} do not consider gas fees.   Gas cost is a flat fee, so it does not depend on the position spread or size of transaction. If the activity of the LP does not affect the pool and if the fees collected scale with the wealth that the LP deposits in the pool, then the LP should consider an initial amount of $X$ and $Y$ that would yield  enough fees to cover the flat gas fees. An estimate of the average gas cost gives an estimate of the minimum amount of initial wealth for a self-financing strategy to be profitable. Recall that, at any point in time $t,$ the LP withdraws her liquidity, adjusts her holdings, and then deposits new liquidity. In the data we consider, the average gas fee is $30.7\,$ USD to  provide liquidity, $24.5\,$ USD to  withdraw liquidity, and $29.6\,$ USD to take liquidity. Average gas costs are obtained from blockchain data which record the gas used for every transaction, and from historical gas prices. The LP pays a flat fee of  $84.8$ USD per operation when implementing the strategy in the pool we consider, so the LP strategy is profitable, on average, if the initial wealth deposited in the pool is greater than $1.8 \times 10^6$ USD. 

Fee income of the LP strategy is limited by the volume of liquidity taking activity in the pool, so one should not only consider increasing the initial wealth to make the strategy more profitable. There are $4.7$ LT transactions per minute and the average volume of LT transactions is $477,275$ USD per minute, so if the LP were to provide $100\%$ of the liquidity, the average fee income per operation would be 1,431 USD.

{
\section{Discussion: modelling assumptions \label{sec:modelling_assumptions}}

This section summarises our modelling assumptions and discusses their implications, strengths, and weaknesses.


\paragraph{Continuous trading.} Our model assumes continuous repositioning of the LP's position. However, when interacting with  blockchains, updates occur at the block validation frequency. For instance, the Ethereum network's blocks are validated every 13 seconds, on average. Moreover, within each block, the transactions form a (random) queue that determines the priority with  which they are executed. {Finally, price formation in Ethereum blockchains leads to sandwich attacks; see \cite{capponi2023paradox} for more details.} Our model can be extended to include delays inherent to blockchains and we refer the reader to the work in \cite{cartea2021shadow} and \cite{cartea2021latency}.

\paragraph{Rebalancing.} Continuous repositioning requires rebalancing the LP's assets which incurs costs as discussed in Section \ref{sec:costs}. To model this aspect, we assume that the LP pays costs that are proportional to the quantity of asset $Y$ held in the pool. In practice, the exact costs depend on variations in the holdings between two consecutive liquidity positions. Thus,  liquidity provision strategies should balance large variations in the holdings with fee revenue, PL, and concentration cost. However, the nonlinearity in the CL constant product formula complicates the mathematical modelling of this aspect of trading costs. 
Moreover, rebalancing costs depend on the cost structure of the trading venue where rebalancing trades are executed. 

\paragraph{Gas fees.} Our model assumes that gas fees paid by the LP to interact with the blockchain are flat and constant. In practice, gas fees are stochastic and depend on exogenous factors such as the price of electricity and network congestion. These could be included in our model by considering a stochastic gas fee that is observed by the LP. {Also, our model is optimal when the spread of liquidity position is continuously adjusted to account for changes in the stochastic drift and profitability of liquidity taking according to \eqref{eq:optimalspeed}. Thus, profitable liquidity provision using our strategy requires a large initial wealth to overcome gas fees from continuous trading. However, we expect the strategy to remain profitable in discrete time when the stochastic drift $\mu$ and the stochastic profitability $\pi$ remain stable, so the LP only rebalances her liquidity position when either the drift $\mu$ or the pool fee rate $\pi$ undergo large changes. }

\paragraph{Concentration costs.} The specific microstructure of CL markets features a new type of investment risk which we refer to as concentration risk. To capture the losses due to concentration risk in a continuous-time framework, we introduced an instantaneous cost which is inversely proportional to the spread of the LP's position, and we showed that it captures the losses due to concentration risk accurately. In practice, LPs must tailor the estimation of the concentration cost parameter $\gamma$ in \eqref{eq:feedyn_0} to the rebalancing frequency and to the volatility of the marginal rate $Z.$

\paragraph{Asymmetry.} Our model assumes a fixed relation between the asymmetry of the LP's position and the drift in the marginal rate. This relation fits observed data but also leads the LP to skew the position to capture more LT trades and to profit from expected rate movements. Future work will consider a richer characterization of the asymmetry because it may be desirable for LPs to adjust the asymmetry of their position as a function of other state variables or as a controlled variable

\paragraph{Fee dynamics.} 
Our model assumes that the distribution of fee revenue $\pi$   among LPs is stochastic and proportional to the size of the pool to reflect that large pools attract more trading flow because trading is cheaper. We also make the simplifying assumption that fees are uncorrelated to the price. In practice, the dynamics of fee revenue may be correlated to those of the volatility of the rate, which is also related to the concentration cost parameter $\gamma.$ Future work will consider more complex relations among these variables. 

\paragraph{Profitability condition.} We impose specific dynamics for the fee rate $\pi$ such that it satisfies a profitability condition \eqref{assumption:1} that allows us to obtain an admissible strategy. While  we use this constraint to solve the optimal liquidity provision problem, it also represents an adequate and natural measure for LPs to assess the profitability of liquidity provision in different pools before depositing their assets.

\paragraph{Constant volatility.} At present, CFMs with CL mainly serve as trading venues for crypto-assets which are better described with a diffusion model with stochastic volatility. It is straightforward to extend our strategy to this type of models.

\paragraph{Market impact.} Finally, our analysis does not take into account the impact of liquidity provision on liquidity taking activity, however, we expect liquidity provision in CPMs with CL to be profitable in pools where the volatility of the marginal rate is low, where liquidity taking activity is high, and when the gas fee cost to interact with the liquidity pools is low. These conditions ensure that the fees paid to LPs in the pool, adjusted by gas fees and concentration cost, exceed PL so liquidity provision is viable.

\subsection{Discussion: related work}

Closest to this work are the strategic liquidity provision models proposed in \cite{fan2021strategic} and \cite{fan2022differential} which also consider CL markets. Both models allow LPs to compute liquidity positions over different intervals centered around the marginal rate within a given time horizon;   \cite{fan2021strategic} only consider static LP strategies which do not use of reallocations, and  \cite{fan2022differential} reposition liquidity whenever the price is outside of the position range. Both approaches only focus on maximising fee revenue and rely on approximations of the LP's objective and  use  Neural Networks to obtain context-aware approximate strategies. In contrast, our model leads to closed-form formulae that explicitly balance fee revenue with concentration risk, predictable loss, and rebalancing costs, while allowing LPs to use price signals (potentially based on exogenous information) to improve trading performance.

}

\section{Conclusions}

We studied the dynamics of the wealth of an LP in a CPM with CL who implements a self-financing strategy that dynamically adjusts the range of liquidity. The wealth of the LP consists of the position value and fee revenue. We showed that the position value depreciates due to PL and the LP widens her liquidity range to minimise her exposure to PL. On the other hand, the fee revenue is higher for narrow ranges, but narrow ranges also increase concentration risk.

We derived the optimal strategy to provide liquidity in a CPM with CL when the LP maximises expected utility of terminal wealth. This strategy is found in closed-form for log-utility of wealth, and it shows that liquidity provision is subject to a profitability condition. In particular, the potential gains from fees, net of gas fees and concentration costs, must exceed PL. Our model shows that the LP strategically adjusts the spread of her position around the reference exchange rate; the spread depends on various market features including tthe volatility of the rate, the liquidity taking activity in the pool, and the drift of the rate.

\clearpage

\appendix

\section{Uniswap v3 ETH/USDC pool data statistics\label{sec:appx:data}}

ETH represents \textit{Ether}, the Ethereum blockchain native currency. USDC represents \textit{USD coin}, a currency fully backed by U.S. Dollars (USD). The fees paid by LTs is  $0.05\%$ of trade size; the fee is deducted from the quantity paid into the pool by the LT and distributed among LPs; see equation \eqref{eq:remuneration_LP}.

Uniswap v3 pools can be created with different values of the LT trading fee, e.g., $0.01\%,$ $0.05\%,$ $0.30\%,$ or $1\%,$ called fee tiers. Additionally, different pools with the same asset pair may coexist if they have different fee tiers. Once a pool is created, its fee tier does not change.

\begin{table}[h]
\centering
\begin{footnotesize}
\begin{tabular}{c || r  r} 
\hline 
 & LT  & LP \\ [0.5ex]
\hline
Number of instructions &  2,654,347 & 68,434 \\ [0.5ex]
Average daily number &  & \\ [-0.5ex] 
of instructions &  4,720 & 471 \\ 
\hline
Total USD volume & $\approx$ \$  262$\times 10^9$	 & $\approx$ \$ 232 $\times 10^9$ \\ 
Average daily USD volume & \$ 554,624,500	 & \$ 863,285 \\ 
\hline
Average LT transaction &  & \\[-0.5ex] 
or LP operation size & \$ 98,624 & \$ 3,611,197 \\ [0.5ex] 
Average interaction frequency &  13 seconds &  590 seconds   \\ [0.5ex]
\hline
\hline 
\end{tabular}
\end{footnotesize}

\caption {LT and LP activity in the ETH/USDC pool between 5 May 2021 and 18 August 2022: Total and average daily count of LT transactions and LP operations in the pool, total and average daily size of LT transactions and LP operations in the pool in USD, average LT transaction size and average LP operation size in USD dollars, and average liquidity taking and provision  frequency. }
\label{table:datadescr}
\hfill
\end{table}

\section{Proofs}

\subsection{Proof of Proposition \ref{prop:1} \label{sec:proofs:hjb1}}

To solve the problem \eqref{eq:valuefunc}, we introduce an equivalent control problem. First, define the process $\left(\tilde \pi_t\right)_{t\in[0,T]} = \left( \pi_t - \eta_t\right)_{t\in[0,T]}$ with dynamics $$d\tilde{\pi}_{t}=\Gamma\,\left(\overline{\pi}-\tilde{\pi}_{t}\right)\diff t+\psi\,\sqrt{\tilde{\pi}_{t}}\,\diff B_{t}\,,$$ where $\tilde \pi_0 = \pi_0 - \eta_0$ and $\eta$ is in \eqref{eq:eta stochastic}. 

We introduce the performance criterion $\tilde u^{\delta}\colon[0,T] \times \R^4 \rightarrow \R$ given by
\begin{align}
\label{eq:perfcriteria_mu_tilde}
    \tilde u^{\delta}(t,\tilde{x},z, \pi,\mu)=\mathbbm{E}_{t,\tilde{x},z, \tilde \pi,\mu}\left[\log\left(\tilde{x}_{T}^{\delta}\right)\right]\, ,
\end{align}
and the value function $\tilde u:[0,T] \times \R^4 \rightarrow \R$ given by
\begin{equation}
\label{eq:valuefunc_mu_tilde}
    \tilde u(t,\tilde{x},z,\tilde \pi,\mu)=\underset{\delta\in\mathcal{A}}{\sup} \, u^{\delta}(t,\tilde{x},z,\pi,\mu)\,.
\end{equation}
Clearly, the problems \eqref{eq:valuefunc_mu_tilde} and \eqref{eq:valuefunc} are equivalent, and the value functions satisfy $u(t,\tilde{x},z,\pi,\mu) = \tilde u(t,\tilde{x},z,\tilde \pi,\mu)$ for all $\left(t,\tilde x, z, \pi, \mu\right)\in[0,T] \times \R^4$ and for all $\tilde \pi = \pi - \eta \in \R$, where $\eta = \frac{\sigma^{2}}{8}-\frac{\mu}{4}\left(\mu-\frac{\sigma^{2}}{2}\right)+\frac{\varepsilon}{4}\,.$

The value function in \eqref{eq:valuefunc_mu_tilde} admits the dynamic programming principle, so it satisfies the HJB equation
\begin{align}
\label{eq:HJB_mu}
0=&\,\partial_{t}w+\frac{1}{2}\sigma^{2}\,z^{2}\,\partial_{zz}w+\mu\,Z\,\partial_{z}w+\Gamma\left(\overline{\pi}-\tilde{\pi}\right)\partial_{\tilde{\pi}}w+\frac{1}{2}\,\psi^{2}\,\tilde{\pi}\,\partial_{\tilde{\pi}\tilde{\pi}}w+\mathcal{L}^{\mu}w\\
&
+\underset{\delta\in\mathbbm{R}^{+}}{\sup}\Bigg(\frac{1}{\delta}\left(4\,\tilde{\pi}+4\,\eta-\frac{\sigma^{2}}{2}\right)\tilde{x}\,\partial_{\tilde{x}}w+\mu\,\rho\left(\delta,\mu\right)\,\tilde{x}\,\partial_{\tilde{x}}w+\frac{1}{2}\,\sigma^{2}\,\rho\left(\delta,\mu\right)^{2}\,\tilde{x}^{2}\,\partial_{\tilde{x}\tilde{x}}w
\\
&\qquad\quad\quad\quad-\frac{\gamma}{\delta^{2}}\,\tilde{x}\,\partial_{\tilde{x}}w+\sigma^{2}\,\rho\left(\delta,\mu\right)\,\tilde{x}\,z\,\partial_{\tilde{x}z}w\Bigg)\,,
\end{align}
with terminal condition 
\begin{align}
\label{eq:tchjb_mu}
w(T,\tilde{x},z,\tilde{\pi},\mu)=\log\left(\tilde{x}\right), \quad \forall \left(\tilde{x},z,\tilde{\pi},\mu\right) \in \R^4\,,
\end{align}
where $\mathcal L^\mu$ is the infinitesimal generator of $\mu$.

To study the HJB in \eqref{eq:HJB_mu}, use the ansatz
\begin{align}
\label{eq:ansatz1}
w\left(t,\tilde{x},z,\tilde{\pi},\mu\right)=\log\left(\tilde{x}\right)+\theta\left(t,z,\tilde{\pi},\mu\right)\,,
\end{align}
to obtain the HJB
\begin{align}
\label{eq:HJBtheta_mu}
0=&\,\partial_{t}\theta+\frac{1}{2}\sigma^{2}\,z^{2}\,\partial_{zz}\theta+\mu\,Z\,\partial_{z}\theta+\Gamma\left(\overline{\pi}-\tilde{\pi}\right)\partial_{\tilde{\pi}}w+\frac{1}{2}\,\psi^{2}\,\tilde{\pi}\,\partial_{\tilde{\pi}\tilde{\pi}}w+\frac{\mu}{2}-\frac{1}{8}\,\sigma^{2}
\\&+\mathcal{L}^{\mu}\theta+\underset{\delta\in\mathbbm{R}^{+}}{\sup}\Bigg(\frac{1}{\delta}\left(4\,\tilde{\pi}+4\,\eta-\frac{\sigma^{2}}{2}\right)+\frac{\mu^{2}}{\delta}-\frac{1}{2}\,\sigma^{2}\,\left(\frac{\mu^{2}}{\delta^{2}}+\frac{\mu}{\delta}\right)-\frac{\gamma}{\delta^{2}}\Bigg)\,,
\end{align}
with terminal condition 
\begin{align}
\label{eq:hjbthetatc_mu}
\theta\left(T, z, \tilde \pi, \mu\right) = 0\,, \quad \forall\, (z, \tilde \pi, \mu)\in\R^3\,.
\end{align}

The supremum in the HJB \eqref{eq:HJBtheta_mu} is attained at 
$$\delta^{\star}=\frac{2\,\gamma+\mu^{2}\,\sigma^{2}}{4\,\tilde{\pi}+4\,\eta-\frac{\sigma^{2}}{2}+\mu\left(\mu-\frac{\sigma^{2}}{2}\right)}=\frac{2\,\gamma+\mu^{2}\,\sigma^{2}}{4\,\tilde{\pi}+\varepsilon}\,.$$ 
Thus, \eqref{eq:HJBtheta_mu} becomes
\begin{align}
\label{eq:HJB2_mu}
0=&\,\partial_{t}\theta+\frac{1}{2}\sigma^{2}\,z^{2}\,\partial_{zz}\theta+\mu\,Z\,\partial_{z}\theta+\Gamma\left(\overline{\pi}-\tilde{\pi}\right)\partial_{\tilde{\pi}}\theta+\frac{1}{2}\,\psi^{2}\,\tilde{\pi}\,\partial_{\tilde{\pi}\tilde{\pi}}\theta+\frac{\mu}{2}-\frac{\sigma^{2}}{8}+\mathcal{L}^{\mu}\theta+\frac{1}{2}\frac{\left(4\,\tilde{\pi}+\varepsilon\right)^{2}}{2\,\gamma+\mu^{2}\,\sigma^{2}}\,.
\end{align}

Next, substitute the ansatz
\begin{align}
\label{eq:ansatz2_mu}
\theta\left(t,z,\tilde{\pi},\mu\right)=&\,A\left(t,\mu\right)z^{2}+B\left(t,\mu\right)\tilde{\pi}\,z+C\left(t,\mu\right)\tilde{\pi}^{2}\\&+D\left(t,\mu\right)z+E\left(t,\mu\right)\tilde{\pi}+F\left(t,\mu\right)\,,
\end{align}
in \eqref{eq:HJBtheta_mu}, collect the terms in $Z$ and $\tilde \pi$, and write the following system of PDEs:
\begin{align*}
\begin{cases}
\left(\partial_{t}+\mathcal{L}^{\mu}\right)A\left(t,\mu\right)= & -\sigma^{2}\,A\left(t,\mu\right)-2\,\mu\,A\left(t,\mu\right)\:,\\
\left(\partial_{t}+\mathcal{L}^{\mu}\right)B\left(t,\mu\right)= & -\mu\,B\left(t,\mu\right)+\Gamma B\left(t,\mu\right)\:,\\
\left(\partial_{t}+\mathcal{L}^{\mu}\right)C\left(t,\mu\right)= & 2\,C\left(t,\mu\right)\Gamma-\frac{8}{2\,\gamma+\mu^{2}\,\sigma^{2}}\:,\\
\left(\partial_{t}+\mathcal{L}^{\mu}\right)D\left(t,\mu\right)= & -\mu\,D\left(t,\mu\right)-\Gamma\,\overline{\pi}\,B\left(t,\mu\right)\:,\\
\left(\partial_{t}+\mathcal{L}^{\mu}\right)E\left(t,\mu\right)= & -2\,\Gamma\,\overline{\pi}\,C\left(t,\mu\right)-\psi^{2}\,C\left(t,\mu\right)+\Gamma\,E\left(t,\mu\right)-\frac{4\,\varepsilon}{2\,\gamma+\sigma^{2}\,\mu^{2}}\:,\\
\left(\partial_{t}+\mathcal{L}^{\mu}\right)F\left(t,\mu\right)= & -\Gamma\,\overline{\pi}\,E\left(t,\mu\right)+\psi^{2}\,\eta\,C\left(t,\mu\right)-\frac{1}{2}\frac{\varepsilon^{2}}{2\,\gamma+\sigma^{2}\,\mu^{2}}-\frac{\mu}{2}+\frac{\sigma^{2}}{8}\:,
\end{cases}
\end{align*}
with terminal conditions $A(T,\mu)=B(T,\mu)=C(T,\mu)=D(T,\mu)=E(T,\mu)=F(T,\mu)=0$ for all $\mu \in \R\,.$

First, note that the PDEs in $A$, $B$, and $D$ admit the unique solutions $A = B = D = 0\,.$ Next, we solve the PDE in $C\,.$ Use Itô's lemma to write
$$C\left(T,\mu_{T}\right)=C\left(t,\mu_{t}\right)+\int_{t}^{T}\left(\partial_{t}+\mathcal{L}^{\mu}\right)C\left(s,\mu_{s}\right)\diff s\,.$$
Next, replace $\left(\partial_{t}+\mathcal{L}^{\mu}\right)C\left(s,\mu_{s}\right)$ with $2\,C\left(s,\mu_{s}\right)\Gamma-\frac{8}{2\,\gamma+\mu_{s}^{2}\,\sigma^{2}}$ to obtain $$C\left(T,\mu_{T}\right)=C\left(t,\mu_{t}\right)+2\,\Gamma\int_{t}^{T}C\left(s,\mu_{s}\right)\diff s-\int_{t}^{T}\frac{8}{2\,\gamma+\mu_{s}^{2}\,\sigma^{2}}\,\diff s\,.$$
Take expectations to get the equation $$C\left(t,\mu_{t}\right)=	\E_{t,\mu}\left[-2\,\Gamma\int_{t}^{T}C\left(s,\mu_{s}\right)\diff s+\int_{t}^{T}\frac{8}{2\,\gamma+\mu_{s}^{2}\,\sigma^{2}}\,\diff s\right]\,.$$
Now consider the candidate solution function $$\hat{C}\left(t,\mu_{t}\right)=\E_{t,\mu}\left[\,\int_{t}^{T}\frac{8}{2\,\gamma+\mu_{s}^{2}\,\sigma^{2}}\exp\left(-2\,\Gamma\left(s-t\right)\right)\,\diff s\right]$$ and write \begin{align*}
    &\E_{t,\mu}\left[-2\,\Gamma\int_{t}^{T}\hat{C}\left(s,\mu_{s}\right)\diff s+\int_{t}^{T}\frac{8}{2\,\gamma+\mu_{s}^{2}\,\sigma^{2}}\,\diff s\right]\\=&\E_{t,\mu}\left[-2\,\Gamma\int_{t}^{T}\E_{s,\mu}\left[\,\int_{s}^{T}\frac{8}{2\,\gamma+\mu_{u}^{2}\,\sigma^{2}}\exp\left(-2\,\Gamma\left(u-s\right)\right)\,\diff u\right]\diff s+\int_{t}^{T}\frac{8}{2\,\gamma+\mu_{s}^{2}\,\sigma^{2}}\,\diff s\right]\\=&\E_{t,\mu}\left[\,\int_{t}^{T}\frac{8}{2\,\gamma+\mu_{s}^{2}\,\sigma^{2}}\exp\left(-2\,\Gamma\left(s-t\right)\right)\,\diff s\right]\,.
\end{align*}
Thus, $\hat C$ is a solution to the equation in $C$ and by uniqueness of solutions, we conclude that $C = \hat C\,.$

Follow the same steps as above to obtain the solution $$E\left(t,\mu\right)=\E_{t,\mu}\left[\,\int_{t}^{T}\left(\left(2\,\Gamma\,\overline{\pi}+\psi^{2}\right)C\left(s,\mu\right)+\frac{4\,\varepsilon}{2\,\gamma+\sigma^{2}\,\mu_{s}^{2}}\right)\exp\left(-\Gamma\left(s-t\right)\right)\,\diff s\right]$$ to the PDE in $E$, and the solution $$F\left(t,\mu\right)=\E_{t,\mu}\left[\,\int_{t}^{T}\left(\Gamma\,\overline{\pi}\,E\left(s,\mu_{s}\right)+\psi^{2}\,\eta_s\,C\left(s,\mu_{s}\right)-\frac{1}{2}\frac{\varepsilon^{2}}{2\,\gamma+\sigma^{2}\,\mu_{s}^{2}}-\frac{\mu_{s}}{2}+\frac{\sigma^{2}}{8}\right)\diff s\right]\,$$ to the PDE in $F\,,$ where $\eta_s = \frac{\sigma^{2}}{8}-\frac{\mu_s}{4}\left(\mu_s-\frac{\sigma^{2}}{2}\right)+\frac{\varepsilon}{4}\,,$ which proves the result.  \qed

\subsection{Proof of Theorem \ref{thm:verif} \label{sec:proofs:hjb2} }
Proposition \ref{prop:1} provides a classical solution to \eqref{eq:HJB_mu}. Therefore, standard results apply and  showing that \eqref{eq:optimalspeed} is an admissible control is enough to prove that \eqref{eq:hjbsol} is the value function \eqref{eq:valuefunc}. Specifically, use the form of the optimal control $\delta^{\star}$ in \eqref{eq:optimalspeed} {to obtain
\begin{equation}
    0<\frac{1}{\delta_{s}^{\star}}=\frac{\tilde\pi_s+\varepsilon}{\sigma^{2}\mu_s^{2}+2\,\gamma}\leq\frac{\tilde\pi_s+\varepsilon}{2\,\gamma}\,,\qquad \forall s\in[t,T]\,,
\end{equation}
where $\tilde \pi_s = \pi_s - \eta_s,$ thus $\delta^{\star}$ is an admissible control.} \qed

\bibliographystyle{elsarticle-harv}
\bibliography{references}

\end{document}